\documentclass[aps,prd,twocolumn,showpacs,groupedaddress,floatfix]{revtex4}

\usepackage{graphicx}  
\usepackage{dcolumn}   
\usepackage{bm}        
\usepackage{amssymb}   
\usepackage{multirow}

\newcommand{\met}       {\mbox{$\not\!\!E_T$}}

\begin{document}

\hspace{5.2in}\mbox{FERMILAB-PUB-07/143-E}
\title{Measurement of the $t\bar{t}$ production cross section in
  $p\bar{p}$ collisions using dilepton events}


%
\author{V.M.~Abazov$^{35}$}
\author{B.~Abbott$^{75}$}
\author{M.~Abolins$^{65}$}
\author{B.S.~Acharya$^{28}$}
\author{M.~Adams$^{51}$}
\author{T.~Adams$^{49}$}
\author{E.~Aguilo$^{5}$}
\author{S.H.~Ahn$^{30}$}
\author{M.~Ahsan$^{59}$}
\author{G.D.~Alexeev$^{35}$}
% ...

\affiliation{\vspace{0.1 in}(The D\O\ Collaboration)\vspace{0.1 in}}
% ...
\affiliation{$^{5}$University of Alberta, Edmonton, Alberta, Canada,
                Simon Fraser University, Burnaby, British Columbia, Canada,
                York University, Toronto, Ontario, Canada, and
                McGill University, Montreal, Quebec, Canada}
% ...
\affiliation{$^{28}$Tata Institute of Fundamental Research, Mumbai, India}
% ...
\affiliation{$^{30}$Korea Detector Laboratory, Korea University, Seoul, Korea}
% ...
\affiliation{$^{35}$Joint Institute for Nuclear Research, Dubna, Russia}
% ...
\affiliation{$^{49}$Florida State University, Tallahassee, Florida 32306, USA}
% ...
\affiliation{$^{51}$University of Illinois at Chicago,
                Chicago, Illinois 60607, USA}
% ...
\affiliation{$^{59}$Kansas State University, Manhattan, Kansas 66506, USA}
% ...
\affiliation{$^{65}$Michigan State University,
                East Lansing, Michigan 48824, USA}
% ...
\affiliation{$^{75}$University of Oklahoma, Norman, Oklahoma 73019, USA}
% ...

\date{June 4, 2007}

\begin{abstract}
We present a measurement of the $t\bar{t}$ pair production cross section 
in $p\bar{p}$ collisions at \( \sqrt{s}=1.96 \) TeV utilizing 
approximately 425 pb$^{-1}$ of data collected with the D0 detector.  
We consider decay channels containing two high $p_{T}$ charged
leptons (either $e$ or $\mu$) from leptonic decays of both top-daughter $W$
bosons.  These were gathered using four sets of selection criteria,
three of which required that a pair of fully identified leptons 
({\em i.e.}, $e\mu$, $ee$, or $\mu\mu$) be found.  The fourth approach
imposed less restrictive criteria on one of the lepton candidates and
required that at least one hadronic jet in each event be tagged as
containing a $b$ quark.  For a top quark mass of 175 GeV, the
measured cross section is 7.4 $\pm$1.4\thinspace(stat) 
$\pm$1.0\thinspace(syst) pb and for the current Tevatron average
top quark mass of 170.9 GeV, the resulting value of the 
cross section is 7.8 $\pm$1.8\thinspace(stat+syst) pb.
\end{abstract}


\maketitle 

\section{Introduction}
\subsection{The Top Quark}
\label{sec:top_quark}
The top quark, first observed by the CDF and D0 collaborations in
1995~\cite{top_discovery_cdf,top_discovery_d0}, is the 
heaviest elementary particle so far observed. Its mass is sufficient to
allow decay to hypothesized particles such as the charged Higgs and to probe
electroweak symmetry breaking physics. At the Fermilab Tevatron
Collider, top quark
production occurs predominantly in top-antitop quark ($t\bar{t}$) pairs. For
a center of mass energy of $\sqrt{s}=1.96$~TeV, leading order QCD
suggests that $t\bar{t}$ production results from quark-antiquark
annihilation about 85$\%$ of the time, while gluon-gluon fusion is
responsible for the remaining 15$\%$~\cite{Yao:2006px}. Recent theoretical
calculations predict, for an assumed top quark mass ($m_t$) of 175~GeV,
an inclusive top quark pair production cross section at $\sqrt{s}=1.96$~TeV of
6.7 pb with an uncertainty of less than
15$\%$~\cite{Cacciari:2003fi,Kidonakis:2003qe}.  If the observed
production cross section were to differ significantly from the standard
model prediction, it would be evidence of new physics, such as exotic
top quark decays or new production mechanisms such as $t\bar{t}$
resonances~\cite{Hill:1993hs}.    Significant deviation among measured
cross sections obtained from the observations 
of different top quark
decay channels would also indicate the presence of new physics.
It is therefore important to precisely measure the top quark pair
production cross section using each possible final state.
Previous measurements by the CDF and D0 
experiments~\cite{Yao:2006px,CDF:Run2,Abazov:2005ey,Abazov:2006ka} show 
good agreement with the theoretical expectation within uncertainties.
The most precise cross section measurement reported by D0 is 
6.6$\pm$1.0 pb~\cite{Abazov:2006ka} in the lepton+jets final state 
and using secondary vertex tagging algorithm to identify $b$ jets.

\subsection{Top Quark Decays and the Dilepton Signature} 
\label{sec:decay_signature}
According to the standard model, the top quark decays almost 100$\%$ of
the time to a $W$ boson and a $b$ quark. For approximately 6$\%$ of
$t\bar{t}$ pairs, both $W$ bosons decay leptonically to generate a final
state containing a pair of electrons, a pair of muons, or an electron
and a muon~\cite{Yao:2006px}.  This produces a unique event signature
consisting of two high transverse momentum ($p_T$) charged leptons,
significant missing transverse energy ($\met$) from the associated
neutrinos, and two high $p_T$ jets from the $b$ quarks. 

Despite low branching ratios relative to
channels with hadronic $W$ boson decays, the dilepton channels are
advantageous for study because few standard model background processes
have two high $p_T$ leptons and neutrinos in their final states. Those
which do usually do not contain two high $p_T$ jets. For example,
electroweak diboson production can result in two isolated, high $p_T$
leptons and neutrinos, but suffers from a low cross section and can be
discriminated against by requiring high $p_T$ jets. Drell-Yan production
of $(Z/\gamma^*)+$jets events has no direct decay process to dilepton
final states with real neutrinos.  $(Z/\gamma^*)$ decay to $\tau$
particles produces neutrinos but suffers from a low branching ratio and
a softer lepton $p_T$ spectrum relative to top quark events.

\subsection{The Content of this Article}
This paper describes a new measurement of top quark decays to final
states containing a pair of electrons or muons, or one electron and one
muon. Section~\ref{sec:apparatus} contains a description of the
experimental setup used to collect the data used for the measurement.
A discussion of the Monte Carlo samples that aided our interpretation of
this data is in Sec.~\ref{sec:simulation}.
Section~\ref{sec:triggers} includes a description of
the triggering system used to acquire the data, and
Sec.~\ref{sec:reco} contains descriptions of the offline
reconstruction techniques used to compute the physical quantities
critical to the extraction of the top quark
signal.  Discussion of the methods used to identify each dilepton decay
mode in the data sample is in Sec.~\ref{sec:analyses}.  Finally, the
computation of the top quark pair production cross section is described
in Sec.~\ref{sec:xsec} and the result is summarized in
Sec.~\ref{sec:conclusion}.

\section{Experimental Apparatus}
\label{sec:apparatus}
\subsection{The Fermilab Tevatron Collider's Run II}
\label{sec:tevatron}
The Fermilab Tevatron Collider, a proton anti-proton accelerator,
collided beams at a center-of-mass energy of 1.8 TeV during the period of
operation (Run I) between 1992 and 1996. The D0 detector,
one of two multipurpose detectors designed to study the high energy  
collisions at Fermilab, collected approximately 120 pb$^{-1}$
of data during Run I~\cite{Abachi:1993em}. After significant
improvements to both the accelerator and the D0 detector, Run II began in
March 2001 with the collider operating at a center-of-mass energy of
1.96 TeV.  The increased energy brought an increase in the top
quark pair production cross section of $\approx$30\%.  The analyses
discussed in this paper are based on approximately 425 pb$^{-1}$ of data
collected by D0 between April 2002 and August 2004.  
D0 has performed a similar measurement 
using $\sim$230 pb$^{-1}$ of data~\cite{Abazov:2005ey}.

\subsection{The D0 Detector}
\label{sec:dzero}
This section presents an overview of the experimental apparatus,
emphasizing the subsystems most relevant to the $t\bar{t}$ production
cross section measurement. A more complete description of the upgraded
experiment can be found in Ref.~\cite{Abazov:2005pn}.

The D0 detector comprises three major subsystems which together
identify and measure the energy or momentum of electrons, jets, muons,
and (indirectly) neutrinos --- all of which can be found in the final
states of $t\bar{t}$ decays. The subsystems are the central tracking
detectors, a uranium/liquid-argon calorimeter, and a muon
spectrometer.

The spatial coordinates of the D0 detector are defined as follows: the
positive $z$ direction is along the direction of the proton motion while
positive $y$ is defined as upward with respect to the detector's center,
which serves as the origin. The polar angle $\theta$ is measured with
respect to the positive $z$ direction and the azimuthal angle $\phi$ is
measured with respect to the positive $x$ direction.  The radial
distance $r$ is the perpendicular displacement from the $z$ axis.  The
polar direction is more usually described by the pseudorapidity, defined
as $\eta \equiv -\ln(\tan \theta/2)$.

The central tracking detectors consist of a silicon microstrip tracker
(SMT) and a central scintillating fiber tracker (CFT) located within a 2 T
solenoidal magnetic field.  Together these detectors are responsible for
locating the position of the hard scatter and for measuring the
trajectories and momenta of charged particles.  The SMT can also locate
displaced, secondary vertices which aid in heavy quark tagging.  It is
composed of high-resistivity silicon sensors arranged in barrels and
disks to maximize the detector surface area perpendicular to charged
particle trajectories.  The barrel detectors provide tracking
information at central values of $|\eta|$ ($<1.5$), while the disks
extend coverage out to $|\eta| \approx$ 3.0.  The CFT is constructed from
scintillating fibers mounted on eight concentric support cylinders.  Each
cylinder supports an axial layer of fibers oriented along $z$ and a
stereo layer oriented at a slight angle with respect to $z$.  The
outermost cylinder provides coverage for $|\eta|<1.7$.


The liquid-argon calorimeter surrounds the central tracking detectors.
In addition to providing energy measurements for electrons, photons, and jets,
it can distinguish showers generated by electrons or photons from those
produced by hadrons.  The calorimeter also plays a critical
role in the measurement of the event-wide transverse energy balance used
to identify 
neutrinos.  The system is composed of three parts:  a central
calorimeter (CC) which provides coverage to $|\eta|\approx$ 1 and
north and south endcap calorimeters (EC) which extend coverage to
$|\eta| \approx$ 4.  Because each calorimeter is housed in its own
cryostat, there is a gap in coverage between each EC and the CC, the region
defined by $1.0 < |\eta| < 1.4$. To partially compensate for this, an
intercryostat detector (ICD) made of a series of scintillating tiles is
located between the CC and EC cryostats.

Each calorimeter section has three subsections: an inner electromagnetic
(EM) section which uses thin uranium absorber plates, a fine hadronic
section which uses uranium-niobium alloy plates, and a coarse hadronic
section which uses copper or stainless steel absorber plates in the CC
or EC, respectively.  The calorimeters are transversely divided into
{\em projective} towers, so-called because the rays along which
the calorimeter cells are oriented project outward from the interaction
center.  Each tower layer is further divided into segments of size
$\Delta\eta\times\Delta\phi$ $=$ 0.1 $\times$ $2\pi/$64, except for the
third layer of the EM section which is segmented twice as finely to
allow for more precise measurement of the EM shower centroid.

The muon spectrometer surrounds the calorimeter cryostats and uses a
combination of wire chambers and scintillation counters to obtain
precise muon spatial and timing information, respectively.  Like the
calorimeter, the muon spectrometer consists of three separate
subdetectors. The central detector covers approximately $|\eta| 
< 1.0$  and the forward systems extend to $|\eta| \approx 2$. Each system
contains three layers of instrumentation, and a 1.8 T iron toroidal magnet is
located between the innermost and second layers.  Each layer contains both wire
chambers and scintillation counters.  The scintillators have response
times sufficiently fast to allow for both muon triggering and out-of-time
background rejection. 

The wire chambers in the central region are proportional drift tubes
(PDTs) oriented to provide maximum resolution for measuring muon bending
angles produced by the toroidal magnetic field.  The innermost central
scintillation counters are segmented in 4.5$^{\circ}$ increments in
$\phi$ to match the CFT segmentation.  Each layer of the forward
spectrometers contains several strata of mini drift tubes (MDTs) and a
set of scintillation counters referred to as {\em pixels}.  The pixels
are projectively arranged from the interaction point with a $\phi$
segmentation of 4.5$^{\circ}$ and an $\eta$ segmentation of
$\approx$0.12.

The luminosity measurement is based on the rate of 
inelastic $p\bar{p}$ collisions 
observed by the luminosity monitors (LM) 
mounted in front of the EC cryostats at $z=\pm 140~\rm{cm}$. The LM consists of two arrays of 
24 plastic scintillator 
counters with photomultiplier readout, and covers the 
$|\eta|$ range between $2.7$ and $4.4$.
The uncertainty on the luminosity measurement is currently estimated 
to be $\pm$ 6.1\% ~\cite{Andeen:2007}.

\section{Event Simulation}
\label{sec:simulation}
Selection efficiencies for $t\bar{t}$ signal events and background survival
rates for each of the analyses were computed using Monte Carlo
simulations of each of the physics processes contributing to the
observed event yields.  
This section provides some details regarding the generation of
the Monte Carlo samples used.  

Simulation began with initial parton generation.  In general, this was
achieved using the {\sc alpgen}~\cite{alpgen} generator,
which contains the exact leading-order (LO) matrix elements for the
processes discussed in the following sections.  Unless otherwise
specified, output from {\sc alpgen} was then convoluted with the
CTEQ5L~\cite{cteq5l} parton distribution functions (PDFs).  Parton showering
was carried out using {\sc pythia}~\cite{pythia}.  Decays of $B$ mesons
were simulated with {\sc evtgen}~\cite{evtgen} and $\tau$-lepton decays
were simulated with {\sc tauola}~\cite{tauola}.

After the modeling of quark and gluon hadronization and unstable
particle decays, the list of generated objects was passed through 
a {\sc geant}-based~\cite{geant} model of the D0 detector.
This provides a detailed simulation of the effects of detector
composition and geometry.  Resolutions for momenta and energies of
leptons and jets, as well as efficiencies for their identification,
were determined in data and compared to their counterparts in Monte
Carlo. Observed discrepancies were used to correct the simulated
samples.

\subsection{$\mathbf{t\bar{t}}$ Production}
\label{sec:ttbar_mc}
Detector acceptance, object reconstruction efficiencies, and the effects
of kinematical cuts were estimated with a sample of simulated
$t\bar{t}\to\ell\ell +X$ decays, where $\ell$ = $e$, $\mu$, or $\tau$.
Seven samples were generated with the
following values of top quark mass ($m_t$):  140, 160, 175,
190, and 210 GeV.  These were used to parameterize 
the signal acceptances as functions of $m_t$. The central
value of the cross section was computed for $m_t=175$ GeV.

\subsection{$\mathbf{(Z/\gamma^*)}$+jets Processes}
\label{sec:zgamma_mc}
The largest background to the $t\bar{t}$ dilepton signal arises from Drell-Yan
$Z/\gamma^*$ production and leptonic decay with associated production
of one or more jets.  To aid in the estimation of these backgrounds, we
generated $Z/\gamma^*\to\ell\ell$ events with one or two partons.  For
each lepton flavor we generated three $M_{\ell\ell}$ regions:  15--60
GeV, 60--130 GeV, and $>$130 GeV. The relative weights of the three
samples were determined from the ratios of their LO cross sections.

The absolute normalizations of these background samples were set by the
number of $Z/\gamma^*$+jets $\to \ell\ell$+jets events observed in
control samples selected from data.  These were chosen by requiring that
the reconstructed dilepton mass be near the $Z$ boson mass so that the samples
were rejected by the signal selection criteria described in
Secs.~\ref{sec:ee} and \ref{sec:mumu}.
This normalization was carried out at an early stage of event selection
where the $t\bar{t}$ yield is a negligible fraction of the selected
sample.  The efficiency of further kinematical selections for
$Z/\gamma^*$ events was then derived from the details of the
simulated samples.


\subsection{Diboson+jets Production}
\label{sec:diboson_mc}
$WW\to\ell\ell +X$ and $WZ\to\ell\ell +X$ (where $\ell$ = $e$ or $\mu$)
production in association with one or two jets contributes to a lesser 
degree to the selected samples. As for all the other Monte Carlo samples
used, we generated these processes at LO with {\sc alpgen}, but the
impact of PDFs was simulated using CTEQ4L~\cite{Lai:1996mg}.  
The resulting samples were normalized using the ratio of NLO to LO
diboson production cross sections calculated without explicit jet
requirements.~\cite{dibosonxs}.

\section{Triggering}
\label{sec:triggers}
The analyses described in this paper made use of data
collected by triggering on the presence of objects consistent with the
dilepton signature:  electrons, muons, central tracks, and jets.
Correlations between these objects and event-wide variables like \met,
though available at all trigger levels, were not utilized in data
collection.  This section begin with a brief description of the D0
trigger system and then provides a description of the triggering
conditions used to collect the dilepton samples analyzed.  A more
detailed discussion of the triggering system is available in
Ref.~\cite{Abazov:2005pn}.

\subsection{The D0 Triggering System}
The D0 triggering system consists of three separate levels, each of which
examines successively fewer events in ever greater
detail.  The first stage (Level 1) is a collection
of custom hardware triggers that 
accepts data from all the major detector subsystems at a rate of 1.7 MHz
and generates an acceptance rate of around or below 2 kHz.  In the second stage
(Level 2), microprocessors associated with each detector subsystem
reconstruct physics objects which are passed on to a global processor
that generates decisions based upon all the objects in an event.
Level 2 provides a maximum accept rate of around 1 kHz.  The final
trigger stage (Level 3), applies more sophisticated algorithms to
data from precision readout of the detector components to further
reduce the overall acceptance rate to around 50 Hz.  Events passing Level 3
are written to tape.

At Level 1, the muon trigger searches for patterns of scintillator and
wire chamber hits consistent with muons traversing the multiple layers
of the muon detector.  {\em Loose} Level 1 muons are constructed
from scintillator hits only, while {\em tight} muons include
corresponding patterns of hits in wire chambers~\footnote{Note that these
muon quality criteria differ from those discussed in
Sec.~\ref{sec:muID}.}.  
Additionally, some Level 1 muon triggers
require that a matching track be found by the central track trigger
(CTT).  CTT tracks are found by analyzing patterns of axial CFT hits.
All eight axial layers must register a hit, and the curvature of the
resulting patterns provides a $p_T$ estimate that is used for a
threshold requirement.

At Level 2, the muon-finding algorithm uses more precise timing
information to improve the quality of muon candidates.  In general, a
combination of wire and scintillator hits both inside and outside the
toroid iron is required. Level 3 uses tracks found in the central
tracker to identify the most probable position for the hard scatter.
This position, also called the primary vertex, is used to refine
momentum estimates from reconstructed
muon-track bending by the toroidal field, and the result is used to
apply momentum threshold requirements. Additionally, Level 3 is capable
of reconstructing central tracks with hits missing and its algorithms
make use of CFT stereo information.
 
The Level 1 calorimeter trigger inputs are electromagnetic (EM) and
hadronic (H) trigger tower energies summed over a transverse area of
$\Delta\eta\times\Delta\phi=$ 0.2 $\times$ 0.2.  Electron
candidates only include energy collected in the EM section of the
calorimeter, while jet candidate energies include the H towers. 
At Level 2, calorimeter objects are reconstructed from trigger towers
using the Level 1 objects as seeds.  The Level 2 jet algorithm clusters
5 $\times$ 5 groups of towers centered on the seed towers.  Electron
candidates are formed by clustering each EM seed with the highest $E_T$
neighboring tower.  

The Level 3 calorimeter triggers use the precision readout chain and
the reconstructed primary vertex position to improve energy and position
resolution relative to Level 2.  Jets and electrons are formed using a
simple cone algorithm~\cite{Blazey:2000qt}.  {\em Loose} Level 3
electrons must have most of their energy deposited in the EM layers and
they must meet basic shower shape criteria.  {\em Tight} Level 3 electrons
must survive additional shape criteria.  Additional background
suppression is also achieved in some triggers by requiring that a
matching central track be found.

\subsection{Dilepton Triggers}
The triggers used to collect the dilepton samples required
that the lepton signatures distinguishing each channel were present at
multiple triggering levels.  In order to reduce rate but maintain
overall trigger efficiency for a given channel, a logical OR of multiple
triggers having different conditions tightened in a complementary manner
was sometimes used.   
Brief summaries of the trigger conditions used for each analysis channel
are presented here, and more detailed breakdowns of the requirements are
available in Appendix~\ref{sec:trig_appendix}.

The $e\mu$ triggers required that an electron with an $E_T$ of at least
5 GeV and a loose muon were found at Level 1.  In some cases a Level 2
muon was also required, but otherwise the remaining conditions involved
electrons reconstructed at Level 3.  A loose Level 3 electron with
$E_T>$ 10 GeV was always included in the requirements, and at higher
luminosity the energy threshold was increased and this requirement was
combined in an OR with a tight electron with $E_T>5$ GeV.

The dielectron triggers usually included the requirement that two
electrons, each with $E_T>$ 6 GeV, were found at Level 1.  In some cases
only one electron with $E_T>$ 11 GeV was required at Level 1.  A Level 2
requirement was only included for later periods containing high luminosity
conditions.  It required that two electrons, each with $E_T>$ 18 GeV, be
found.  The Level 3 condition always included at least one loose
electron with $E_T>$ 10 GeV.  For later data taking periods, a second
electron was added to the Level 3 condition, and the energy threshold
and quality requirements were tightened.

To maximize efficiency, the $\mu\mu$ channel made use of high $p_T$
single muon triggers and switched to dimuon trigger requirements when
the single muon triggers were prescaled due to high rates.  The single
muon triggers required a tight muon at Level 1, while the dimuon
triggers used loose Level 1 muons.  All triggers used for the dimuon
channel demanded that one muon be found at Level 2, sometimes with a
$p_T>$ 3 GeV requirement.  The Level 3 requirements also involved single
muon signatures, but complementary conditions were sometimes combined in
a logical OR.  These signatures included a Level 3 muon with a $p_T$ of
at least 6 GeV or a Level 3 central track with a $p_T$ of at least 5
GeV.

Since the $\ell$+track channels did not require that two identified
leptons be found in each candidate event, they relied on high $p_T$
single lepton triggers.  In some cases these triggers included jet
requirements.  At Level 1, the $e$+track triggers demanded the presence of
either one EM object with $E_T>$ 10 GeV or two EM objects, each with
$E_T>$ 3 GeV.  In some cases the single electron condition was
coupled with the requirement that two Level 1 jets were found, each with
$E_T>$ 5 GeV.  The Level 2 conditions included the presence of one
electron with $E_T>$ 10 GeV.  Some triggers also asked that
two jets, each with $E_T>$ 10 GeV, be found at Level 2.  Level 3
requirements included at least one electron with $E_T>$ 15 GeV.  Some
triggers also required that two jets, each with $E_T>$ 15 GeV, be
found at Level 3.

The $\mu$+track triggers required that at least one loose Level 1 muon,
sometimes with a matching central track, be found.  Some triggers also
demanded that at least one jet with $E_T>$ 3 GeV be found.  The Level 2
conditions usually required that one muon be found and sometimes
also required one jet with $E_T>$ 8 GeV to be present.  Level 3
conditions alternately included 1 jet with $E_T>$ 10 GeV, one muon with
$p_T>$ 15 GeV, or one central track with $p_T>$ 10 GeV.  Sometimes the
track and muon requirements were combined in a logical OR.

The efficiencies of trigger conditions on single objects were estimated
using data samples selected to remove triggering bias.  The efficiency
for a $t\bar{t}$ event to satisfy a trigger condition was then
estimated by folding per-muon, per-electron, and per-jet efficiencies into
Monte Carlo simulated events.  A similar process was used for those
background estimations that are based upon simulation.

Trigger terms related to electrons, muons, and tracks were analyzed using 
reconstructed $Z$ boson decays to dilepton final states.  In each such decay,
one lepton was matched to triggering and reconstruction requirements so
that the other lepton could be used for unbiased efficiency
measurements.  This method, known as ``{\em tag and probe},'' was used
to perform most of the high $p_T$ lepton efficiency measurements used in
the dilepton analyses.  Hadronic jet triggers were studied using events
passing either muon-based or electron-based triggers.
Electron-triggered events were required to have exactly one electron
that was both matched to electron trigger objects at all levels and
separate from the jet considered in the efficiency measurement.  

Efficiency measurements 
were parameterized in terms of the kinematic variables
$p_T$, $\eta$, and $\phi$ of offline reconstructed objects.
Uncertainties in these parameterizations were derived from fits to the
observed variable distributions.

Separate efficiencies were estimated for Level 1 ($L1$), Level 2 ($L2$), and
Level 3 ($L3$) conditions and the total event probability $P(L1,L2,L3)$
was estimated as
\begin{equation}
  P(L1,L2,L3) = P(L1)\cdot P(L2|L1) \cdot P(L3|L1,L2),
\end{equation}
where $P(L2|L1)$ and $P(L3|L1,L2)$ are the conditional probabilities for
an event to satisfy a set of criteria provided it has already passed
offline selection and previous levels of triggering.  The overall
trigger efficiency for $t\bar{t}$ events was then computed as the
luminosity-weighted average of the event probabilities associated with
each data taking period.

\section{Event Reconstruction}
\label{sec:reco}
\subsection{Track Reconstruction}
\label{sec:tracks}
Charged particle trajectories were reconstructed from the patterns of
energy deposits (or ``hits'') that they left in the tracking detectors.
Track reconstruction at D0 involved two distinct steps:  track finding
and track fitting.   Two complementary track finding algorithms were
used in event reconstruction.  The first is a histogramming approach
based upon the Hough Transform --- a method originally developed for
finding tracks in bubble chambers~\cite{Hough:1959}.  An alternate
track-finding approach began with groups of hits in the SMT barrels.
These were fitted to a track hypothesis and the result was used to form
a road in which to search for hits in additional detector layers.

The candidate track lists resulting from the two track-finding
approaches were combined and passed to a Kalman~\cite{Kalman:1960} track
fitter.  This made use of an interacting propagator which propagates
tracks through the D0 tracking system while taking into account magnetic
curvature and interactions with detector material.  The fitter
incrementally adds hits to tracks using the input candidates to define
roads.  The resulting track fit allows for the calculation of optimal
track parameters, with errors, on any surface. 

The tracking momentum scale was determined by comparing the dimuon invariant
mass distribution for $Z\rightarrow\mu\mu$ decays in data with
expectation from simulation based upon the world average $Z$ boson mass
computed by the Particle Data Group~\cite{Yao:2006px}.
In order for the simulated track momentum resolutions to match those
observed in the data sample, an additional random smearing of track
parameters was performed.

The measured transverse momentum resolution can be expressed as
\begin{equation}
\frac{\sigma(1/p_T)}{1/p_T} = 
\sqrt{\frac{(0.003p_T)^2}{L^4} + \frac{(0.026)^2}{L\cdot \sin\theta}}.
\end{equation}
Here $p_T$ is measured in GeV and $L$ is the {\em normalized track bending
lever arm}.  $L$ is equal to 1 for tracks with $|\eta|<$ 1.62 and is
computed as $\tan\theta$/$\tan\theta '$ otherwise ($\theta '$ is the angle
at which the track exits the tracker).

\subsection{Primary Vertex Identification}
\label{sec:pvtx}
The principal task of primary vertex (PV) finding is to identify tracks
originating from the hard scatter and to separate these from tracks
generated in superimposed minimum bias events.  The algorithm first
reconstructed one or more vertices and then selected the hard scatter
vertex from among them by considering the $p_T$ distribution and number 
of tracks associated with each vertex.

The vertex reconstruction algorithm included three steps:  track
clustering, track selection, and vertex fitting.  First, tracks were
clustered in $z$ by considering their relative separations.  Second,
tracks kept for fitting were required to have at least 2 SMT hits and
$p_T \ge$ 0.5 GeV.  Each track's distance of closest approach in the
$x$-$y$ plane ($d_{CA}$) to the nominal interaction position was also
considered:  the $d_{CA}$ significance ($S=d_{CA}/\sigma_{d_{CA}}$) for
a candidate track had to be less than 3.  Finally, for every $z$
cluster, an iterative vertex search yielded a vertex position.  A
probability that a vertex originated from a minimum-bias interaction was
assigned based on the transverse momenta of its associated tracks. The
vertex with the lowest probability was selected as the {\em primary}, or
hard scatter, vertex.  To further ensure the quality of selected primary
vertex candidates, we required that they be within the SMT fiducial
region ($|z_{PV}|\leq 60$ cm) and that they have at least three
associated tracks.  

In multijet data events, the position resolution of the primary vertex
in the transverse plane is around 35 $\mu${\rm m}, convoluted with a typical
beam spot size of 30 $\mu$m. The vertex resolution in the direction of
the beam line is $\approx$1 mm.

\subsection{Muon Identification}
\label{sec:muID}
Muon identification was based on matches between charged particles
found in the central tracking system and trajectories reconstructed in
the muon systems.  Tracks in the muon detectors comprised straight-line
segments, or {\em stubs}, formed from combinations of 
scintillator and wire chamber hits in a single layer.  Stubs were formed
separately inside and outside the toroid iron and then paired together
(provided their combinations were consistent with expectations for muons
originating in the interaction region).  Pairs were fitted to
trajectories using knowledge of the toroidal magnetic field and the
expected effects of energy loss and multiple scattering.  The resulting
{\em local} muon momenta were used, along with the directions of the
muons at the inner surface of the muon system, to search for consistent
central tracks with which to form a {\em global} match.  Stubs that were
not used in forming local muons were also used in global matches.  
In all cases, the results of the original central track fits were taken
as the best estimates of muons' momenta, since the resolution of the
central tracker is far superior to that of the local muon system.

To reduce the impact of muon detector noise, requirements were made on
the number and location of scintillator and wire chamber hits.  Two sets
of muon quality requirements were used in the analyses discussed in this
paper:  {\em tight} and {\em loose}.  Tight muons were required to have
wire and scintillator hits both inside and outside the toroid iron.  The
loose criteria also accepted muons formed from single stubs with both
types of hits either inside or outside the toroid.


Central tracks pointing into the fiducial volume of the muon
detectors ({\em i.e.}, with $|\eta|<$2) were considered
as candidates for matches to muon tracks.  To ensure that a central
track was well-reconstructed, the $\chi^2$ per degree of freedom
of the Kalman fit used by the central tracking algorithm was required
to be less than 4.  Consistency between candidate tracks and the primary
vertex was ensured by two additional cuts:  the $d_{CA}$ significance of
each track must have been less than 3 and the smallest distance in $z$
between it and the primary vertex (PV) must have been less than 1 cm.  The
quality criteria applied to central tracks and matching local muons are
summarized in Table~\ref{tab:muon_quality}.

\begin{table*}
\caption{\label{tab:muon_quality} The quality criteria applied to
muon candidates.  Variable definitions are provided in the text.}
\begin{ruledtabular}
\begin{tabular}{l l}
Cut Level & Requirements \\
\hline
Loose & Local muon stubs inside and/or outside toriod iron, \\
      & \ \ central track with $\chi^2_{\text{Kalman}}/d.o.f. <$ 4,
      $\sigma_{d_{CA}/d_{CA}} <$ 3, \\
      & \ \ and $\Delta z(\text{track},\text{PV}) <$ 1 cm \\
Tight & {\em Loose} and local muon stubs BOTH inside and outside toroid
      iron \\
\end{tabular}
\end{ruledtabular}
\end{table*}

Muons produced in top quark decays can be distinguished from those
originating in heavy quark or other hadronic decays by a lack of
nearby activity in the tracker and the calorimeter.  This feature
motivated two isolation criteria used to select signal candidate muons.
These involved summing visible energies over a region around the central
track associated with the muon.  One variable was computed by summing
the energies of reconstructed tracks and the other variable was derived
from energy deposited in the calorimeter. For background muons, the size
of either of these sums is correlated with the muon energy, while for
signal it is not.  Therefore, scaling the sums by the $p_T$ of the
candidate muon generates variables that tend to be higher for background
than signal.  A cut on either of these variables translates to an upper
limit on surrounding visible energy that increases with muon $p_T$.
Hence these variables provide more efficient criteria than the visible
energy sums alone.

The track-based variable was computed as
\begin{equation}
\displaystyle
\mathcal{E}_{\text{halo}}^{\text{trk}} = 
\frac{1}{p_T^{\mu}} \cdot \sum_{\Delta \mathcal{R}<0.5}p_T^{\text{trk}}.
\label{eq:muon_trkiso}
\end{equation}
Here $\Delta \mathcal{R}$ was defined for each
muon-track pair as their separation in $\eta$-$\phi$ space,
$\sqrt{\Delta\eta^2 + \Delta\phi^2}$.  The track matched to the
muon was excluded from the sum.  Similarly, the calorimeter-based
isolation was defined as
\begin{equation}
\displaystyle
\mathcal{E}_{\text{halo}}^{\text{cal}} = 
\frac{1}{p_T^{\mu}} \cdot \sum_{0.1<\Delta \mathcal{R}<0.4} E_T^{\text{cell}},
\end{equation}
where the sum was over individual calorimeter cells.  In each analysis
channel that includes muons in its final state, signal candidate muons
were required to have values of both
$\mathcal{E}_{\text{halo}}^{\text{trk}}$ and 
$\mathcal{E}_{\text{halo}}^{\text{cal}}$ that are less than 0.12.  This
requirement was found to reject more than 99\% of muons originating in
hadronic jet decays and to be about 87\% efficient for a muon coming
from a top quark decay.

\subsection{Electron Identification}
\label{sec:eID}
\label{sec:em_reco}
High $p_{T}$ electrons were identified by the presence of localized
energy deposits in the electromagnetic calorimeter.   Electron
reconstruction began with the formation of clusters through the use of
a simple cone algorithm that grouped calorimeter
cells around seed cells having $E_{T}>$ 0.5 GeV.  Any resulting cluster
having $E_{T}>$ 1 GeV was then grouped with all EM towers within a cone
of radius $\Delta \mathcal{R}=0.4$.  The centroid of a cluster was
calculated as the energy weighted mean value of the coordinates of its
cells in the third layer of the EM calorimeter.  Additional quality criteria
were then applied to reject clusters resulting from photons and hadronic
activity.

Electrons (and photons) deposit almost all of their energy in the EM
section of the calorimeter while hadrons typically penetrate into the
hadronic sections. Hence an electron is expected to have a large EM
fraction ($f_{EM}$), which is defined as the ratio of summed energies
deposited in the EM layers to the total energy deposited inside the
clustering cone.

The longitudinal and lateral shower profiles of an EM cluster were
required to be compatible with those of an electron (or photon).  This
was done by forming a $\chi^2_{\text{cal}}$ based on a comparison of the
energy depositions in each layer of the EM calorimeter and the total
energy of the shower to average distributions obtained from Monte Carlo 
simulations.

Electrons produced in top quark decays can be distinguished from those
originating in heavy quark or other hadronic decays by a lack of
nearby activity in the calorimeter.  The electromagnetic isolation
fraction ($f_{\text{iso}}$) was used to quantify the degree of isolation
of an EM cluster and was defined as 
\begin{equation}
f_{\text{iso}}=\frac{E_{\text{tot}}(\Delta
\mathcal{R}<0.4)-E_{EM}(\Delta \mathcal{R}<0.2)} {E_{EM}(\Delta
\mathcal{R}<0.2)},
\end{equation}
where $E_{\text{tot}}(\Delta \mathcal{R}<0.4)$ is the total energy
within a cone of size $\Delta \mathcal{R}=0.4$ around the direction of
the cluster, and $E_{EM}(\Delta \mathcal{R}<0.2)$ is the energy in a
cone of size $\Delta \mathcal{R}=0.2$ summed over EM layers only.  

In order to suppress photons and some hadronic jet backgrounds, an
electron candidate was required to have an associated track in the
central tracking system within $|\Delta\eta_{EM,\text{trk}}|<0.05 \:\:\mathrm{
and }\:\: |\Delta\phi_{EM,\text{trk}}|<0.05$ of the center of the EM cluster.

To further isolate real electrons, an {\em electron likelihood} formed from
seven variables was computed.  These variables included $f_{EM}$,
$\chi^2_{\text{cal}}$, the ratio of calorimeter transverse energy to track
transverse momentum ($E_T^{\text{cal}}/p_T^{\text{trk}}$), the quality
of the spatial matching between the central track and the EM cluster
($\chi^2_{\text{spatial}EM-\text{trk}}$), the $d_{CA}$ of the track to
the primary 
vertex, the number of tracks in a $\Delta R = 0.05$ cone, and the sum of
the transverse momenta of all tracks in a cone of size  $\Delta R = 0.4$
around the EM-associated track. Smoothed, normalized distributions of
each of these variables were made from signal-like ({\em i.e.},
$Z\rightarrow ee$) events and background ({\em i.e.}, QCD dijet) data
samples. For each discriminating variable $x_i$, these distributions
provided probabilities $P^i_{\text{sig}}(x_i)$ and $P^i_{\text{bkg}}(x_i)$ for an EM
object to be from a real and a fake electron, respectively.  The
following likelihood discriminant was used to distinguish between
real electrons and fakes from hadronic objects.
\begin{equation}
\mathcal{L}_{e}({\mathbf x})=\frac{P_{\text{sig}}({\mathbf{x}})}{P_{\text{sig}}
({\mathbf x})+P_{\text{bkg}}({\mathbf x})},
\end{equation}
where ${\mathbf x}$ is the vector of likelihood variables.
The probabilities were formed without regard to correlations between the
likelihood variables, {\em i.e.},
\begin{equation}
P_{\text{sig/bkg}}({\mathbf x}) = \prod_{i=1}^7 P_{\text{sig/bkg}}^{i}(x_i).
\end{equation}

Three classes of electrons were considered on the basis of the
aforementioned quantities: loose (or EM cluster), medium, and
tight.  The criteria applied to each category are listed in
Table~\ref{tab:electron_quality}.  Less than 0.5\% of all hadronic jets
that passed loose EM object identification criteria survived the tight 
cuts.  The efficiency of the medium quality cuts in data was found to
be about 90\% in the CC and about 63\% in the EC.  With respect to the
medium requirements, the additional likelihood cut in the tight
criteria is about 86\% efficient in the CC and 84\% efficient in the
EC.

\begin{table*}
\caption{\label{tab:electron_quality} The quality criteria applied to
electron candidates.  Variable definitions are provided in the text.}
\begin{ruledtabular}
\begin{tabular}{l l}
Cut Level & Requirements \\
\hline
Loose (EM cluster) & $p_{T}^{\text{cluster}}>1.5$ GeV, $f_{EM}>0.9$,
$f_{\text{iso}}<0.20$, and $|\eta|<$ 1.1 OR 1.5 $<|\eta|<$ 2.5 \\ 
Medium & {\em Loose} and $f_{\text{iso}}<0.15$,
$\chi^2_{\text{cal}}<50$, and central track match \\
Tight & {\em Medium} and $\mathcal{L}_{e}>0.85$ \\
\end{tabular}
\end{ruledtabular}
\end{table*}

The $ee$ analysis selected events with two tight electrons and the
$e+track$ analysis used one tight electron.  The $e\mu$ analysis had
smaller backgrounds and therefore applied medium criteria to the signal
electron candidate in each event.  Because of the poor 
energy resolution for electrons reconstructed in the regions between the
CC and EC sections, all electron-based analyses eliminated
candidates having  1.1 $<|\eta|<$ 1.5.  Electrons with $|\eta|>$ 2.5 were
also removed from consideration in order to suppress multiple scattering
backgrounds.

The EM energy scale was established by requiring that the $Z$ boson mass
reconstructed in track-matched dielectron events match the world average
$Z$ boson mass computed by the Particle Data Group~\cite{Yao:2006px}. By
requiring that both electrons in $Z$ candidate events be in either the
CC or the EC, independent absolute energy scale factors were obtained
for each portion of the calorimeter.  These were applied to high $p_{T}$
electrons and photons. The calibration at lower energies was also
checked using $J/\psi\rightarrow ee$ decays.  The energy resolution for
electrons in the CC or EC is $\sigma (E)/E \approx (15 / \sqrt{E} \oplus
4)$\% or $\sigma (E)/E \approx (21 / \sqrt{E} \oplus 4)$\%, respectively
(here $E$ is measured in units of GeV).

\subsection{Jet Identification}
\label{sec:jetID}
Particle jets were reconstructed from energy deposition in the
calorimeter using a seed-based, improved legacy cone
algorithm~\cite{Baur:2000bi} with a cone radius of $\Delta
\mathcal{R}=0.5$.  In this scheme, seeds were formed by clustering
calorimeter cells above an energy threshold of 0.5 GeV.  All resulting
pre-clusters having summed energies above 1.0 GeV were then fed into an
iterative clustering algorithm.  If any of the resulting proto-jets
shared energy, they were either split or merged so that each calorimeter tower
was assigned to at most one reconstructed jet.  Finally, only those jets
having energies of at least 8 GeV were retained for further
consideration.

\label{sec:jet_qual}
Additional quality criteria were applied to clustered jets to suppress
backgrounds originating from noise and other instrumental effects.
Some of these selection cuts were based on variables that discriminate
against particular sources of noise.  The {\em coarse hadronic
fraction} ($f_{CH}$) is the fraction of the total energy in a jet that
is contained in the outer, noisier layers of the calorimeter.  The {\em
hot fraction} ($f_{\text{hot}}$) is the ratio of the energy of the most energetic
cell in a jet to that of its next-to-highest energy cell.  Both
$f_{\text{hot}}$ and $N_{90}$ (the number of cells containing 90\% of the total
energy in a jet) were used to suppress jets clustered around single cells
that fired erroneously.  Since noise generally did not appear simultaneously
in the precision readout chain and in the separate Level 1 trigger readout,
the ratio of the Level 1 energy to the precision readout energy in a jet
($f_{L1\sum E_{T}}$) is another powerful discriminant against jets due
to noise.

Other requirements were made on jet candidates to remove clusters that
don't originate from partons generated in the hard scatter.  The EM
fraction ($f_{EM}$, see Sec.~\ref{sec:eID}), was used to remove
reconstructed electrons and photons.  To eliminate backgrounds from low
energy multiple interactions, far forward candidates were also
eliminated.  The particular values of all jet quality cuts are
summarized in Table~\ref{tab:jet_quality}.

\begin{table*}
\caption{\label{tab:jet_quality} The quality criteria applied to
jet candidates.  Variable definitions are provided in the text.}
\begin{ruledtabular}
\begin{tabular}{l l}
Cut & Target \\
\hline
0.05 $<f_{EM}<$ 0.95 & Noise and EM particles \\
$f_{CH}<0.4$ & Noise in coarse hadronic layers \\
$f_{\text{hot}}<10$ & Jets clustered around single cell \\
$N_{90}>1$ & Jets clustered around single cell \\
$f_{L1\sum E_{T}}>0.4$ ($|\eta|<0.7$) & Noise in readout \\
$f_{L1\sum E_{T}}>0.2$ ($0.7<|\eta|<1.6$) & Noise in readout \\
$|\eta|<2.5$ & Extra soft scattering interactions in an event \\
\end{tabular}
\end{ruledtabular}
\end{table*}

After these initial cuts, some electrons still remained among the
reconstructed jet objects.  In order to avoid the resulting ambiguity, 
jet candidates overlapping medium quality electrons (see
Table~\ref{tab:electron_quality}) within $\Delta \mathcal{R}=0.5$
were considered only as EM objects.

A data-to-Monte-Carlo correction factor that accounts for possible
differences in the jet reconstruction and identification efficiencies was
determined with back-to-back $\gamma$+jet events by requiring 
$E_{T}$ balance between the photon and the jet. An $E_{T}$-dependent
scale factor was then obtained separately for the CC, EC and  ICD
regions and applied to Monte Carlo.  

\label{sec:JES}
A number of effects --- including non-linearities in calorimeter response,
non-instrumented material, and noise --- can cause the measured jet energy
to differ from the original particle-level energy.  Jet energy scale
(JES) corrections were applied to adjust jet energies to the
particle level.  Transverse momentum conservation in samples of
$\gamma$ + jet events was used to calibrate JES corrections in data and
simulation.  A more detailed description of this procedure is
available in Ref.~\cite{Abazov:2006}.  The relative uncertainty on the
jet energy calibration is $\approx$ 7\% for jets with 20 $<p_T<$ 250
GeV. 

Jet momentum resolutions were measured using dijet and $\gamma$+jet
events.  For 50 GeV jets in the CC or EC, the resolution was found to be
$\sigma(p_T)/p_T \approx$ 13\% or $\sigma(p_T)/p_T \approx$ 12\%,
respectively.

\subsection{Missing Transverse Energy}
The presence of one or more neutrinos in an event is indicated by an
imbalance of the visible momentum in the transverse plane. Calculation
of this quantity began with the vector sum of the transverse energies of
all calorimeter cells surviving various noise suppression algorithms,
with the possible exception of cells in the coarse hadronic layers.
These were included only if they are clustered within a reconstructed
jet.  The vector opposite to the result is called the {\em raw missing
transverse energy} (\met$^{\text{raw}}$).

As EM and jet energy scale corrections were applied to calorimeter
objects, \met$^{\text{raw}}$ was adjusted through vector subtraction.
Only jets that pass the quality criteria listed in
Table~\ref{tab:jet_quality} were used for the hadronic part of this
correction. The result is called the {\em calorimeter missing transverse
energy} (\met$^{\text{cal}}$).

Since muons are minimum ionizing particles, they deposit only a small
fraction of their energy in the calorimeter and \met$^{\text{cal}}$ does not
properly account for their presence.  Therefore, the momenta of all
the identified muons of tight quality (see Table~\ref{tab:muon_quality})
in an event were subtracted vectorially from \met$^{\text{cal}}$ after first
deducting the muons' expected energy depositions in the calorimeter. For
the $\mu\mu$ analysis, isolated {\em loose} muons which did not pass
tight quality requirements were also removed from \met$^{\text{cal}}$.  A
similar procedure was followed for the track identified with a signal
muon in the $\mu$+track analysis.  The fully corrected imbalance is
simply called the {\em missing transverse energy} (\met).

\subsection{$b$ Jet Tagging}
\label{sec:btag}
Bottom quark jets were identified using a secondary vertex tagging (SVT)
algorithm that exploits the long lifetime of $b$ hadrons. The algorithm
used is the same as that used in previously published D0\ $t\bar{t}$
production cross section measurements~\cite{Abazov:2005ey,Abazov:2006ka}.

The SVT procedure began by clustering tracks in $z$ into track jets.
Track jets were reconstructed using a $\Delta{\cal{R}}=$ 0.5 cone
algorithm to cluster tracks with $p_T>$ 0.5 GeV, at least two SMT hits,
and $|d_{CA}|<$ 0.15 cm. The $z$-projection of a candidate track's
$d_{CA}$ onto the beam line ($z_{dca}$) was required to be within 0.4 cm
of the $z$ position of the PV.

Within each track jet, tracks having $d_{CA}$ significances 
greater than 3.5, $\chi^2$/d.o.f. less than 3, and transverse momentum 
greater than 1~GeV were paired to form seed vertices.  
Vertices consistent with having come from $\gamma$ conversions or 
$K^0_S$ or $\Lambda$ decays were removed from consideration.  
Additional tracks pointing to a surviving seed were attached to 
it based on their contribution to the vertex $\chi^2$.
Vertices resulting from this process
were selected as secondary vertex (SV) candidates based upon the
collinearity of their component tracks, the $\chi^2$ of their fits, and 
their decay length significances ($L_{xy}/\sigma_{L_{xy}}$).  Here
$L_{xy}$, the decay length, is the distance between the primary and
secondary vertices in the plane transverse to the beam line and
$\sigma_{L_{xy}}$ includes the uncertainty in the primary vertex
position.  The decay length can be positive or negative, depending on
the sign of its projection onto the track jet axis.  Secondary vertices
corresponding to the decay of $b$ and, to some extent,  $c$ hadrons 
are expected to have large positive decay lengths.

A calorimeter jet was tagged as a $b$ jet if a secondary vertex with
$L_{xy}/\sigma_{L_{xy}}>$ 7 was found within $\Delta{\cal{R}}<$ 0.5.
If a jet contained at least one secondary vertex with
$L_{xy}/\sigma_{L_{xy}}<-7$, the jet was labeled {\em negatively
tagged}. Negative tags resulted from
fake or mis-reconstructed tracks, or from the effects of multiple
scattering in detector material. 
Negative tags were used to estimate the probability to misidentify a 
light flavor (a $u$, $d$, or $s$ quark or a gluon) jet as a $b$ jet (the
{\em mis-tagging rate}).
   
The overall event tagging probability for a particular process depends upon
the flavor composition of the jets in the final state and on the event
kinematics.  This probability was estimated through the application of
tagging rates measured in data to each jet in simulation.  A brief
description of tagging probability measurements is given here, and a
more detailed discussion is available in Ref.~\cite{Abazov:2006ka}.

In order to decouple tagging efficiency measurements from detector
geometry effects, jets considered for tagging were required to pass
additional {\em taggability} criteria.  A taggable jet had to have its axis 
matched to within $\Delta{\cal{R}}<$ 0.5  with the axis of a track jet.
The SMT hit requirement for tracks in track jets means that most
taggable jets are associated with PV $z$ positions within $\approx$36 cm
of the center of D0 detector.  Within this range, jets with momenta
above 30 GeV typically have taggabilities of greater than 90\%.

The $b$ tagging efficiency was estimated using a sample of dijet events
enriched in semileptonic decays of bottom and charm hadrons by the
requirement that one jet have an associated muon.  The heavy flavor
content of this sample was further enriched by requiring that the opposite jet
was tagged, either with the SVT algorithm or with a {\em soft lepton}
tag. Soft lepton tagging requires that a muon with a momentum component 
along the direction orthogonal to the jet axis
of at least 0.7 GeV be found within $\Delta{\cal{R}}=$ 0.5.
In order to extract the SVT tagging efficiency, both tagging algorithms
were applied separately and together to the dijet sample.  The resulting $b$
tagging efficiency depends upon both $\eta$ and $p_T$, and a typical 40
GeV taggable jet from top quark decay is tagged about 40\% of the time.  Charm
quark jets have tagging efficiencies around 20\% as large as
those for $b$ quarks.  Light quark mis-tagging probabilities are on the
order of 0.1\%.

\section{Analyses}
\label{sec:analyses}
\subsection{General Considerations}
As discussed in Sec.~\ref{sec:decay_signature}, $t\bar{t}$ decays to
dilepton final states are characterized by the presence of two
high $p_T$ charged leptons, significant \met\ from two neutrinos, and
two or more jets (from $b$ quark fragmentation and initial and final
state radiation).  Analysis of all four channels described here 
therefore begins by
requiring that signatures consistent with two isolated,
oppositely charged leptons and at least two jets be reconstructed. 

The backgrounds at this stage vary with the channel considered,
but generally include Drell-Yan production of $(Z/\gamma^*)+$jets,
diboson production ({\em i.e.}, $WW$ or $WZ$) with jets, and leptonic $W+$jets
events in which another lepton arises from the misidentification of 
one of the jets.  Resonant production of $Z$ bosons which decay into
electron or muon pairs is the dominant background for the $ee$ and
$\mu\mu$ channels, each of which employed cuts based upon \met\ and
the invariant mass of the lepton pair to target this process.

The remaining backgrounds were removed using a combination of kinematic
and topological constraints.  These include a summed transverse energy
$H_T$, defined as
\begin{equation}
\label{eqn:h_t}
\displaystyle
H_T = p_T^{\ell_1} + \sum_{i=1}^2p_T^{j_i};
\end{equation}
where $\ell_1$ denotes the highest $p_T$ lepton, and $i$ in the summation
extends over the two highest $p_T$ jets in the event.  For the $e\mu$
analysis, a cut on $H_T$ was found to be more effective than one on
\met\ in rejecting $Z\to\tau\tau$ background.

Analyses using fully reconstructed leptons ({\em i.e.}, the $e\mu$,
$ee$, and $\mu\mu$ channels) were optimized separately to achieve the
best possible performance using kinematic and topological quantities to
supress dominant backgrounds.  In order to recover some of the
efficiency lost due to lepton identification requirements, an alternate 
approach using one fully reconstructed lepton and one central track was
taken.  To contend with the additional background let in by the lack of
lepton identification requirements on the second lepton, the
$\ell+$track analysis required that at least one jet pass explicit $b$
quark tagging requirements.  The $ee$, $\mu\mu$, and $\ell+$track
selections are not completely orthogonal, and their overlaps
were accounted for in the combined cross section calculation discussed
in Sec.~\ref{sec:xsec}.  The low-background $e\mu$ analysis was designed
to have no overlap with the other channels.

\subsection{The $e\mu$ Channel}
The signature for an $e\mu$ event consists of one high $p_T$ isolated
electron, one high $p_T$ isolated muon, and two or more jets.  The major
backgrounds to this channel come from Drell-Yan production of 
$\tau$ pairs which in turn decay to produce an $e\mu$ pair ({\em i.e.},
$Z/\gamma^* \to \tau\tau \to e \mu +X$) and $WW$ production with jets.
There are additional backgrounds present from misidentified leptons,
particularly electrons.  These are mostly $W$($\to\mu\nu$)+3 jet events
in which one of the jets was misidentified an electron.  Hereafter,
objects misidentified as electrons will also be referred to as {\em
fake} electrons.

Offline selection began with medium electron and tight muon
identification cuts (see Secs.~\ref{sec:eID} and \ref{sec:muID} for
details).  To reject bremsstrahlung events, in which the muon emits a
photon that is mistakenly identified as an electron, the candidate
electron and muon were not allowed to share a common track in the
central tracking detectors.
Additionally, the candidate electron and muon were required to be
matched to tracks of opposite charge.

After this initial lepton selection, the sample was dominated by
background consisting of roughly equal amounts of misidentified leptons
and physics processes leading to legitimate $e\mu$ pairs.  
The backgrounds generally contain jets arising from initial
and final state radiation.  These tend to be softer in $p_T$ than the
$b$ jets that are generated in $t\bar{t}$ decays.  Requiring that two or
more jets (Sec.~\ref{sec:jetID}) be found with $p_T$ of at
least 20 GeV reduces both backgrounds by more than a factor of 50 while
preserving more than two-thirds of the signal.

\begin{figure}
\includegraphics[scale=0.35]{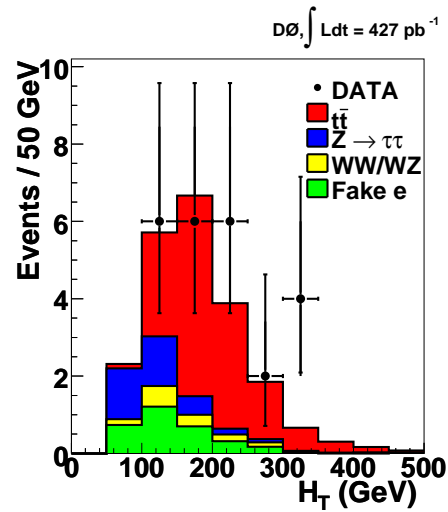}
\caption{\label{fig:emu_HT}  $H_T$ distributions in the $e\mu$ channel
  for expected signal (assuming $m_t=175$ GeV and
  $\sigma_{t\bar{t}}=7$ pb), expected background, and data after both
  lepton and jet requirements (corresponding to the second line
  of Table~\ref{tab:emu_yields}).}  
\end{figure}

In addition to basic lepton and jet identification and energy
requirements, the use of event-wide selection criteria was found to
improve the expected significance of the result.  Several variables were
considered, including \met, $H_T$, and the transverse mass of the
combination of the leptons and the \met.  The performances of cuts on
these quantities, whether alone or in combination, were evaluated using
the expected significance of the background subtracted yield, including
both the statistical uncertainties of the yields and a term reflecting
the dominant jet energy scale systematic uncertainty in the total
background estimate.  It was found that requiring at least 122 GeV of
$H_T$ gave the best performance.  Figure~\ref{fig:emu_HT} shows the $H_T$
distributions for signal and background before this cut and illustrates
its ability to discriminate signal from background.
Table~\ref{tab:emu_yields} shows the impact of the $H_T$ cut on expected
signal and background yields.

\begin{table*}
\caption{\label{tab:emu_yields}Numbers of observed and expected $e\mu$
  events passing the analysis cuts.  The instrumental
  background is from fake electrons.  Expected number of
  $t\bar{t}$ events are for $m_t=175$ GeV and $\sigma_{t\bar{t}}=7$
  pb.  Uncertainties correspond to statistical and systematic
  contributions added in quadrature.} 
\begin{ruledtabular}
\begin{tabular}{l c c c c c}
&      & Total & Instrumental & Physics & \\
& Data & sig+bkg & bkg  & bkg & $t\bar{t}$ \\ 
\hline
Trigger, 1 $e$, $p_T^e>15$ GeV      & & & & \\
\quad + $\ge$1 $\mu$, $p_T^{\mu}>15$ GeV      & & & & \\
\quad + $\ge$2 jets, $p_T^{\text{jet}}>20$ GeV   &  24 &   $22.0^{+3.3}_{-2.9}$ & $3.2^{+2.8}_{-2.0}$ & $4.9^{+1.2}_{-1.4}$& $13.8^{+1.5}_{-1.7}$\\
+ $H_T>122$ GeV                     &  21 &   $17.7^{+2.9}_{-2.4}$ & $2.1^{+2.5}_{-1.7}$ & $2.5^{+0.7}_{-0.7}$& $13.1^{+1.4}_{-1.6}$\\
\end{tabular}
\end{ruledtabular}
\end{table*}

After all cuts are applied, 21 events remain in the data.
Table~\ref{tab:emu_expected} shows the expected background and
signal (assuming $m_t=175$ GeV and $\sigma_{t\bar{t}}=7$ pb)
contributions to the final sample.  

\begin{table}
\caption{\label{tab:emu_expected} 
  A more detailed listing of the expected $e\mu$ signal and background yields presented on the last line of Table~\ref{tab:emu_yields}.
  The expected number of $t\bar{t}$ events is calculated
  assuming $m_t=175$ GeV and $\sigma_{t\bar{t}}=7$ pb.  Uncertainties
  include statistical and systematic contributions added in quadrature.}
\begin{ruledtabular}
\begin{tabular}{l c}
Process $\hspace{20pt}$ Expected number of $e\mu+X$ events & \\
\hline
$t\bar{t}$ (MC)             & $13.09^{+1.42}_{-1.65}$ \\
$Z\to\tau\tau\to e\mu+X$ (MC) & $ 1.46^{+0.38}_{-0.45}$ \\
$WW/WZ \to e\mu+X$ (MC)       & $ 0.99^{+0.40}_{-0.42}$ \\
Fake leptons (data)         & $ 2.14^{+2.50}_{-1.66}$ \\
Total background            & $ 4.58^{+2.56}_{-1.77}$ \\
\end{tabular}
\end{ruledtabular}
\end{table}

\begin{figure}
\includegraphics[scale=0.35]{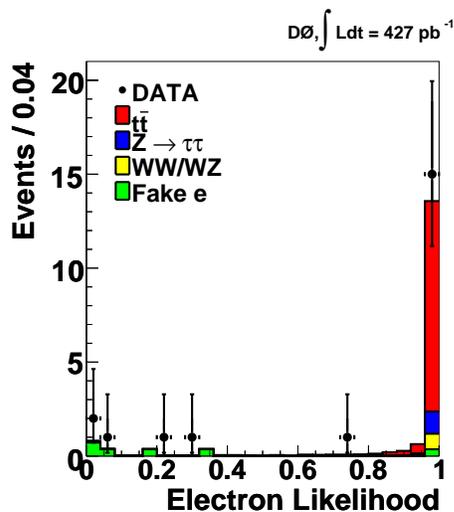}
\caption{\label{fig:emu_likelihood} The electron likelihood distribution
in the $e\mu$ channel.   Data passing all selection criteria are shown
with points.  The fake electron background estimated by a fit
to the data points is shown, along with the simulated likelihood
distributions for the each of the estimated signal and other background
contributions, as shaded histograms.}  
\end{figure}

Contributions from the processes $Z\to\tau\tau\to e\mu +X$ and $WW/WZ \to e\mu +X$
were calculated using the Monte Carlo samples discussed in
Sec.~\ref{sec:simulation}.  The background from fake electrons
was estimated by performing an extended unbinned likelihood fit to the
observed electron likelihood (Sec.~\ref{sec:eID}) distribution in events
passing all selection criteria.  The distribution, which is shown in
Fig.~\ref{fig:emu_likelihood}, was fitted using a likelihood given by 
\begin{equation}
\label{eq:emu_likelihood}
\displaystyle
\mathcal{L}=\prod_{i=1}^N [n_eS(x_i)+n_{\text{fake}}B(x_i)]
\frac{e^{-(n_e+n_{\text{fake}})}}{N!},
\end{equation}
where $i$ is an index that runs over all selected events, $x_i$ is the
corresponding observed value of the electron likelihood, $N$ is the
total number of events, $n_e$ is the number of events with signal-like
electrons, $n_{\text{fake}}$ is the number of events having fake electrons, and 
$S$ and $B$ are the signal and background probability distribution
functions, respectively.  The event counts $n_e$ and $n_{\text{fake}}$
were allowed to float in the fit.

The probability distribution functions used in the fit were determined
with separate data samples enhanced in signal-like and background-like
electrons.  The signal probability distribution function came from a fit
to the likelihood distribution of electrons in oppositely signed
dielectron events selected using standard electron identification cuts
and having low \met.  The background probability distribution function
was determined using events passing all signal selection criteria but
the jet requirements and using an anti-isolation cut on the muon.  The
contribution from signal-like electrons to the resulting sample was
found to be less then 0.5\%.  Note that the fake electron estimate
resulting from Eq.~\ref{eq:emu_likelihood} includes backgrounds
containing both real and fake isolated muons.  The contribution from
events containing legitimately identified electrons and falsely isolated
muons was also investigated and found to be negligible.

\subsection{The $ee$ Channel}
\label{sec:ee}
The signature for an $ee$ event consists of two high $p_T$ isolated
electrons, at least two high $p_T$ jets, and substantial \met.  The
main background to this signature arises from Drell-Yan production of
dielectrons ($Z/\gamma^{*}\rightarrow ee$). Although this process
produces no real \met, mismeasured \met\ can originate from
misreconstructed jet or electron energies or from noise in the
calorimeter.  Other 
backgrounds include Drell-Yan production of $\tau$ pairs which further
decay to dielectrons ($Z/\gamma^*\rightarrow \tau\tau \rightarrow ee +X$)
as well as diboson ($WW/WZ$) production associated with jets. There are
also small backgrounds from  $W(\rightarrow e \nu)+$multijet and QCD
multijet events in which one or two jets are misidentified as isolated
electrons. The background from heavy flavor ($c\overline{c}$,
$b\overline{b}$) production is negligible since electrons from these
decays are typically soft and non-isolated. 

Event selection began with two tight electrons, as described
in Sec.~\ref{sec:eID}, each having $p_{T}>15$ GeV.  The electrons
were also required to have matching central tracks of opposite
charge. This initial selection essentially eliminated any background
from heavy flavor production and significantly reduced the background
from misidentified electrons.  The additional requirement that two jets
be found, each with $p_{T}>20$ GeV, generated a sample dominated by
Drell-Yan $Z/\gamma^{*}$ production with associated jets.
Table~\ref{tab:ee_yields} shows the $ee$ sample composition at this and
subsequent stages of selection.

\begin{table*}
\caption{\label{tab:ee_yields}Numbers of observed and expected $ee$
  events passing the analysis cuts.  The instrumental
  background includes events containing misidentified electrons and
  misreconstructed \met.  Expected number of
  $t\bar{t}$ events are for $m_t=175$ GeV and $\sigma_{t\bar{t}}=7$
  pb.  Uncertainties correspond to statistical and systematic
  contributions added in quadrature.} 
\begin{ruledtabular}
\begin{tabular}{l c c c c c}
&      & Total & Instrumental & Physics & \\
& Data & sig+bkg & bkg  & bkg & $t\bar{t}$ \\ 
\hline
Trigger, $N_e\ge 2$, $p_T^e>15$ GeV     & & & & & \\
\quad + $\ge$2 jets, $p_T^{\text{jet}}>20$ GeV  &  369 &   $428.4^{+79.3}_{-77.1}$  & 
$415.9^{+79.3}_{-77.1}$ & $5.9^{+0.6}_{-1.4}$  & $6.6^{+0.6}_{-0.6}$\\
+ $M_{ee}$ cut  &  88 &   $106.2^{+19.0}_{-23.0}$  & $98.6^{+19.0}_{-23.0}$ & $1.9^{+0.3}_{-0.6}$  & $5.7^{+0.5}_{-0.6}$\\
+ \met\ cut  &  5 &   $5.7^{+0.5}_{-0.6}$  & $0.7\pm0.2$ & $0.7^{+0.2}_{-0.3}$  & $4.3^{+0.4}_{-0.5}$\\
+ sphericity cut  &  5 &   $5.2^{+0.5}_{-0.5}$  & $0.5\pm0.2$ & $0.6^{+0.2}_{-0.2}$  & $4.0^{+0.4}_{-0.5}$\\
\end{tabular}
\end{ruledtabular}
\end{table*}

\begin{figure*}
\includegraphics[scale=0.35]{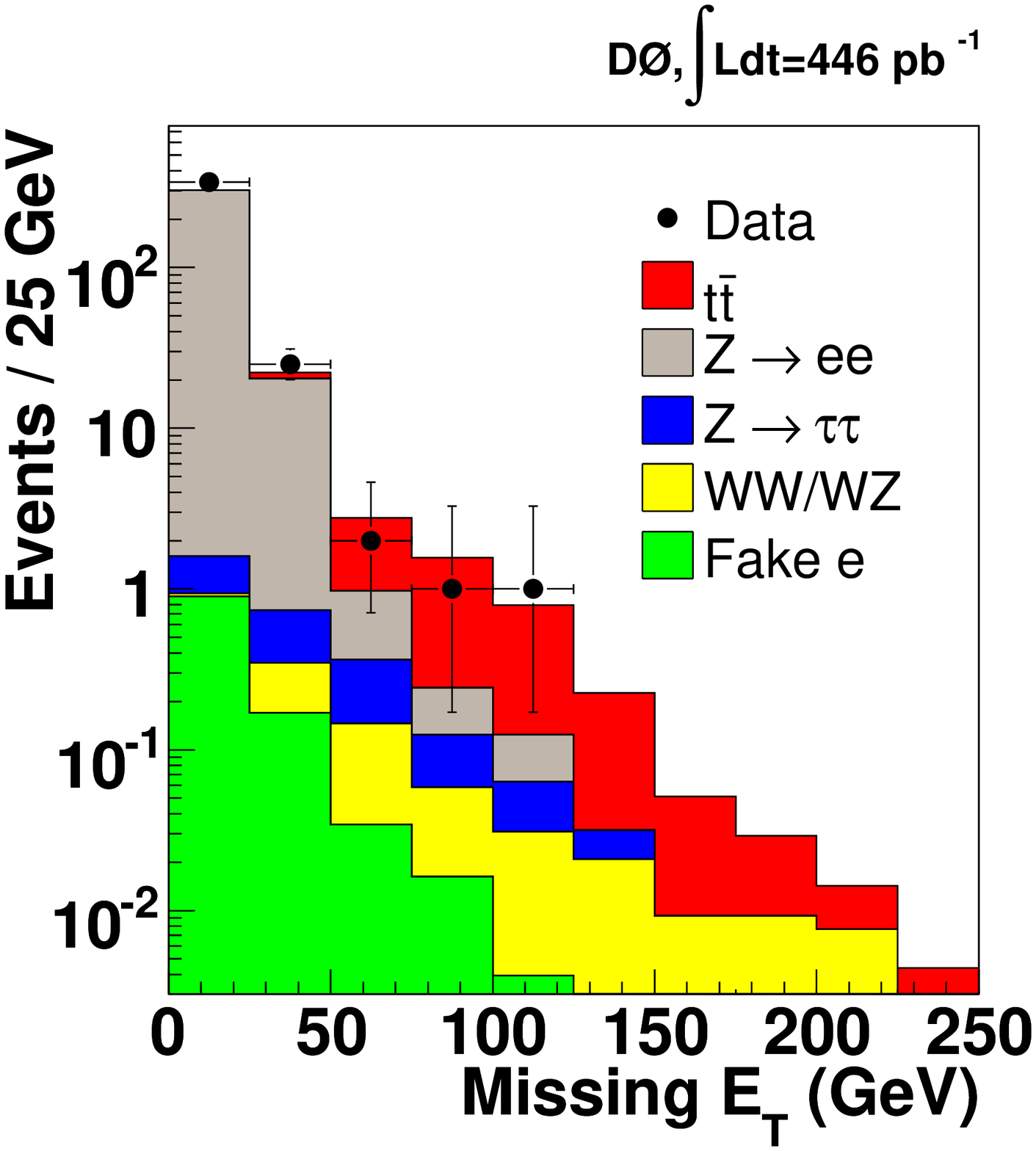}
\includegraphics[scale=0.35]{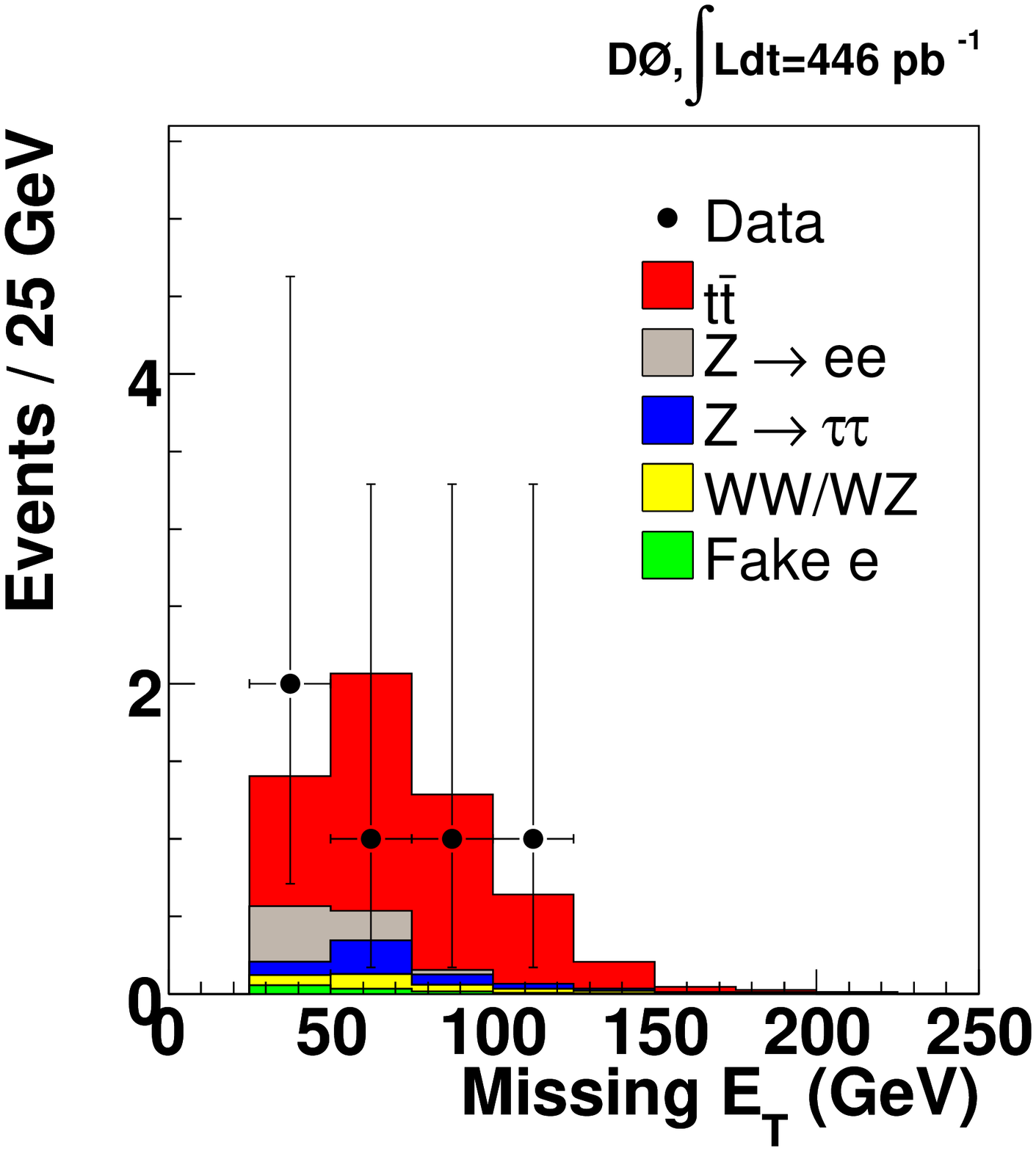}
\caption{\label{fig:met_ee} Expected and observed \met\ distributions
in dielectron events before (left) and after (right) cutting on $M_{ee}$
and \met.}
\end{figure*}

Simultaneous cuts on the \met\ and the dielectron invariant mass
($M_{ee}$) provide a powerful way to suppress most of the
$Z/\gamma^*\rightarrow ee$ background.  We vetoed events with $M_{ee}$
values near the $Z$ boson mass ({\em i.e.}, $80$ GeV $<M_{ee}<$ 100 GeV)
and required $\met > 35$ GeV ($\met > 40$ GeV) for $M_{ee}> 100$ GeV
($M_{ee}< 80$ GeV).  These requirements effectively suppressed the
$Z/\gamma^*\rightarrow ee$ background and brought other backgrounds to a
manageable level while preserving significant signal efficiency.  Their
effect is illustrated in Fig.~\ref{fig:met_ee}.

Finally, a cut on the sphericity of the event was applied in order to
take advantage of the topological peculiarities of $t\bar{t}$ events and
gain even more discrimination between signal and background.
Sphericity ($\mathcal{S}$) is defined as
\begin{equation}
\displaystyle
\mathcal{S}=\frac{3}{2}(\epsilon_1 + \epsilon_2),
\end{equation}
where $\epsilon_1$ and $\epsilon_2$ are the two leading eigenvalues 
of the event-normalized momentum tensor~\cite{Barger:1993ww}. The
tensor ($\mathcal{M}_{xy}$) is calculated as
\begin{equation}
\mathcal{M}_{xy} = \frac{\sum_{i}p_x^ip_y^i}{\sum_{i}(p^i)^2},
\end{equation}
where the index $i$ runs over the leading two electrons and the leading
two jets in the event.

\begin{figure}
\includegraphics[scale=0.35]{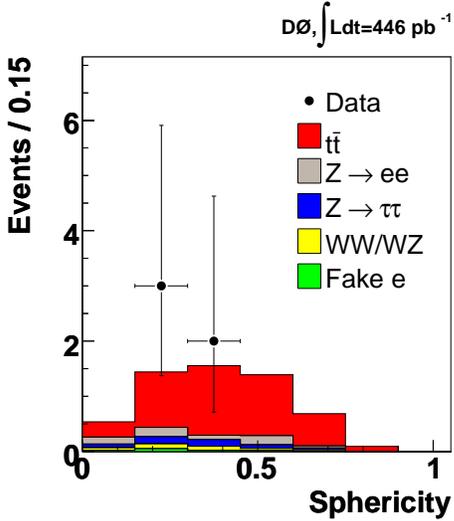}
\caption{\label{fig:sph_ee} The observed and predicted sphericity
distributions of dielectron events after the $M_{ee}$ and \met\ cuts.} 
\end{figure}

Sphericity can take values between 0 and 1.  The applied cut of
$\mathcal{S}>0.15$ rejects events in which jets are produced in a planar
geometry due to gluon radiation and provides a reasonable reduction in
most of the backgrounds. The cut value was chosen using a figure of
merit related to the expected statistical significance of the background
subtracted signal.  The observed and expected sphericity distributions
for events passing the $M_{ee}$ and \met\ cuts are shown in
Fig.~\ref{fig:sph_ee}.

After all cuts, five events remain in the data.
Table~\ref{tab:ee_expected} shows the the expected background and  
signal (assuming $m_t=175$ GeV and $\sigma_{t\bar{t}}=7$ pb)
contributions to the final sample.  

\begin{table}
\caption{\label{tab:ee_expected} 
  A more detailed listing of the expected $ee$ signal and background yields presented on the last line of Table~\ref{tab:ee_yields}.
  The expected number of $t\bar{t}$ events is calculated
  assuming $m_t=175$ GeV and $\sigma_{t\bar{t}}=7$ pb.  Uncertainties
  include statistical and systematic contributions added in quadrature.}
\begin{ruledtabular}
\begin{tabular}{l c}
Process $\hspace{20pt}$ Expected number of $ee+X$ events &\\
\hline
$t\bar{t}$ (MC)             &  $4.04^{+0.40}_{-0.46}$ \\
$Z\to\tau\tau\to ee+X$ (MC)   &  $0.35^{+0.11}_{-0.15}$ \\
$WW/WZ \to ee+X$ (MC)         &  $0.23^{+0.11}_{-0.16}$ \\
Mismeasured \met\ (data)    &  $0.45\pm0.15$ \\
Fake electron (data)        &  $0.09\pm0.03$ \\
Total background            &  $1.12^{+0.22}_{-0.27}$ \\
\end{tabular}
\end{ruledtabular}
\end{table}

\begin{figure}
\includegraphics[scale=0.35]{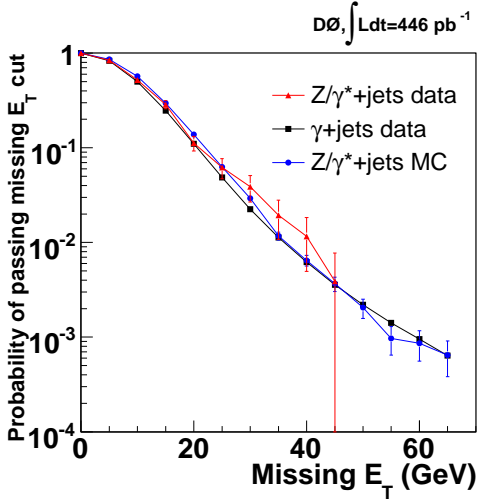}
\caption{\label{fig:met_fake_rate} The probability to pass the $\met$
selection for events with two or more jets in $Z/\gamma^*\rightarrow
e^+e^-$ data, in $\gamma+ 2\; \rm{jets}$ data, and for
$Z/\gamma^*\rightarrow e^+e^-$ Monte Carlo.}
\end{figure}

The background contributions from $Z/\gamma^{*}\rightarrow\tau \tau$
and $WW/WZ(\rightarrow ee+X)$ decays were estimated using the Monte Carlo
samples described in Sec.~\ref{sec:simulation}.  Simulated 
data were also used to estimate contributions from
$Z/\gamma^{*}\rightarrow e e$ decays for cut levels prior to the \met\
requirement in Table~\ref{tab:ee_yields}.  The background due to
mismeasured \met\ was estimated from data using the observed
{\em misreconstruction probability} ($f_{\text{MET}}$) in
$\gamma+$2 jets  
events selected to have kinematics and resolutions similar to the
$Z/\gamma^{*}$ backgrounds. Figure~\ref{fig:met_fake_rate} shows that
the \met\ spectrum in $Z/\gamma^{*}\rightarrow ee$ events with two or more
jets is in good agreement with that observed in the $\gamma+$2 jets sample.
We therefore estimated $f_{\text{MET}}$ as the fraction of
$\gamma+$2 jets events passing the \met\ selection.  

The \met\ misreconstruction probability was multiplied by the number of
events that failed the \met\ cut but passed all other selections.  Thus the
number of expected \met\ background events summed over the low and high
$M_{ee}$ regions is 
\begin{equation}
N_{\text{MET}}^{\text{misreco}} =  N_{M_{ee}<80 \text{GeV}} \times f_{\text{MET}}^{40 \; \text{GeV}} + 
N_{M_{ee}>100 \text{GeV}} \times f_{\text{MET}}^{35 \; \text{GeV}}.
\end{equation}

The background due to electron misidentification was also obtained from
data by calculating the fake electron probability, $f_e$.  This was derived
from a control sample containing two loose EM objects and passing signal
dielectron triggers.  Additional cuts on \met\ and $M_{ee}$ were used to
suppress contributions from signal-like electrons, and the resulting
sample was found to be completely dominated by fake electrons.

The predicted number of fake electrons in the final sample
($N_{e}^{\text{fake}}$) was obtained by multiplying the number of events with
one loose and one tight electron by $f_e$.  At this stage,
$N_{e}^{\text{fake}}$ contains both $W$+jet and QCD multijet backgrounds.  The
latter enters into the sample when two jets mimic the signal electron
signature and misreconstructed \met\ allows the events to pass full
selection criteria.  Since this background was counted along with the
misreconstructed \met\ background obtained from the data, it was removed
from the $N_{e}^{\text{fake}}$ estimate to avoid double counting. This was
achieved by loosening identification cuts on both electron candidates in
the final sample and scaling the yield by the square of $f_e$.  This led
to an estimate of $N_{QCD}$ that was used to correct $N_{e}^{\text{fake}}$.

\subsection{The $\mu\mu$ Channel}
\label{sec:mumu}
The signature for a $\mu\mu$ event consists of two high $p_T$ isolated
muons, two high $p_T$ jets, and large \met.  The
dominant background in the dimuon channel comes from Drell-Yan 
production of muon pairs ($Z/\gamma^* \rightarrow \mu\mu$) in which
misreconstructed objects give rise to mismeasured $\met$. Other backgrounds
include Drell-Yan production of $\tau$ pairs which decay to produce
muon pairs ($Z/\gamma^*\rightarrow\tau\tau \rightarrow \mu\mu +X$) as well
as $WW$ and $WZ$ production with jets. There are also backgrounds from
multijet and $W$+jets events, in which one or more muons from
heavy flavor decay pass isolation criteria.  Hereafter, this class of
muons will also be referred to as {\em fake} isolated muons.

\begin{table*}
\caption{\label{tab:mumu_yields}Numbers of observed and expected $\mu\mu$
  events passing the analysis cuts.  The instrumental
  background includes events containing fake isolated muons and
  misreconstructed \met.  Expected number of $t\bar{t}$ events are for
  $m_t=175$ GeV and $\sigma_{t\bar{t}}=7$ pb.  Uncertainties correspond
  to statistical and systematic contributions added in quadrature.} 
\begin{ruledtabular}
\begin{tabular}{l c c c c c}
&      & Total & Instrumental & Physics & \\
& Data & sig+bkg & bkg  & bkg & $t\bar{t}$ \\ 
\hline
Trigger, $N_{\mu}\ge 2$, $p_T^\mu>15$ GeV & & & & \\
\quad + $N_{\text{jets}}\ge 2$$p_T^{\text{jet}}>20$ GeV & 387 & $382.8^{+23.9}_{-23.5}$ & $371.6^{+23.6}_{-23.2}$ & $3.7^{+0.9}_{-0.9}$  & $7.6^{+0.7}_{-0.7}$\\
+ $\Delta\phi(\mu_{\text{leading}},\met)$--\met\ cut & 5 & $6.2^{+0.8}_{-1.0}$  & $1.7^{+0.3}_{-0.5}$ & $0.5^{+0.2}_{-0.1}$ & $4.0^{+0.4}_{-0.5}$\\
+ $\chi^2 > 4$ & 2 & $3.6^{+0.5}_{-0.5}$ & $0.3^{+0.1}_{-0.2}$  & $0.4^{+0.1}_{-0.1}$ & $3.0^{+0.3}_{-0.4}$ \\
\end{tabular}
\end{ruledtabular}
\end{table*}

\begin{figure*}
\includegraphics[scale=0.35]{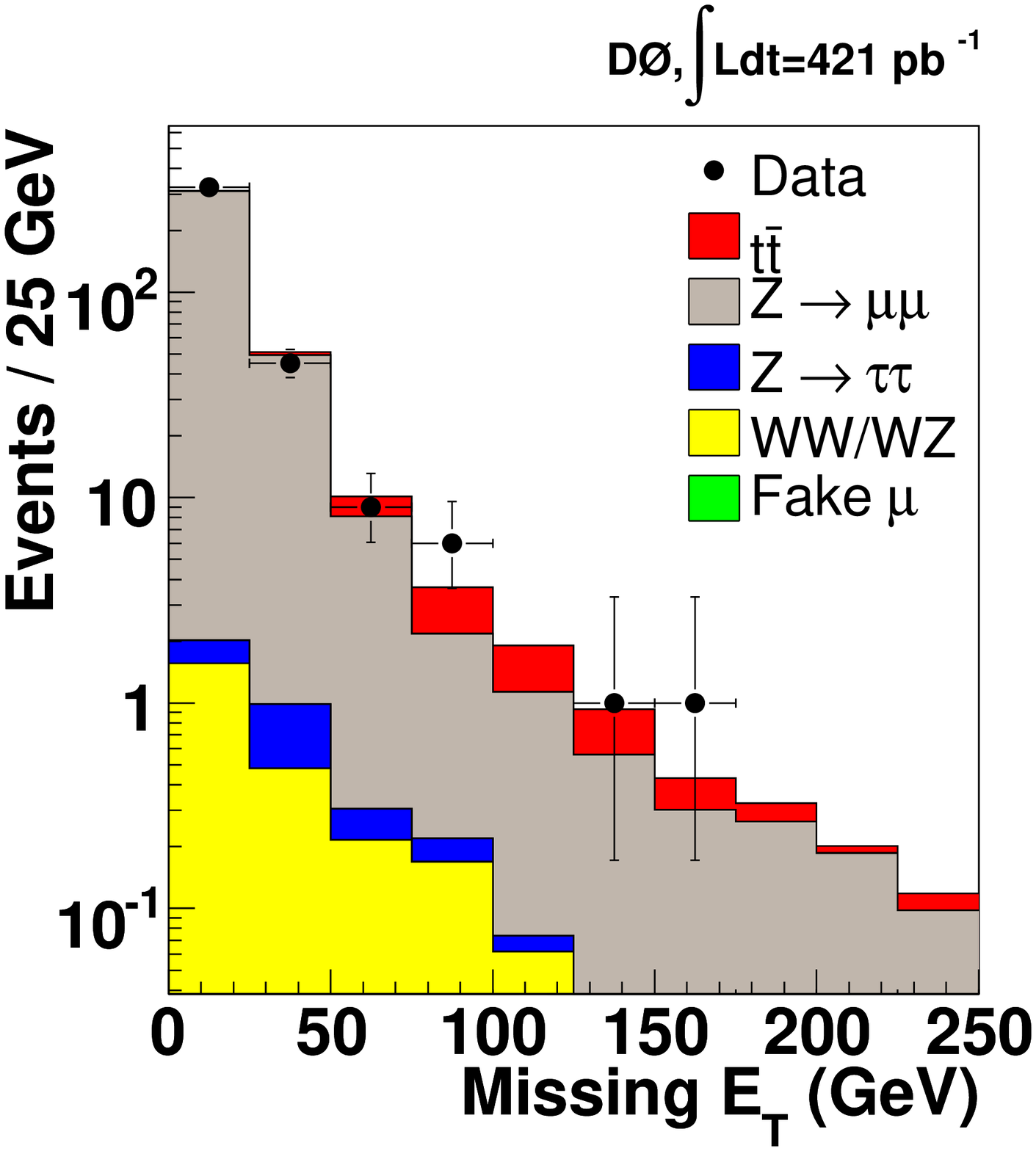}
\includegraphics[scale=0.35]{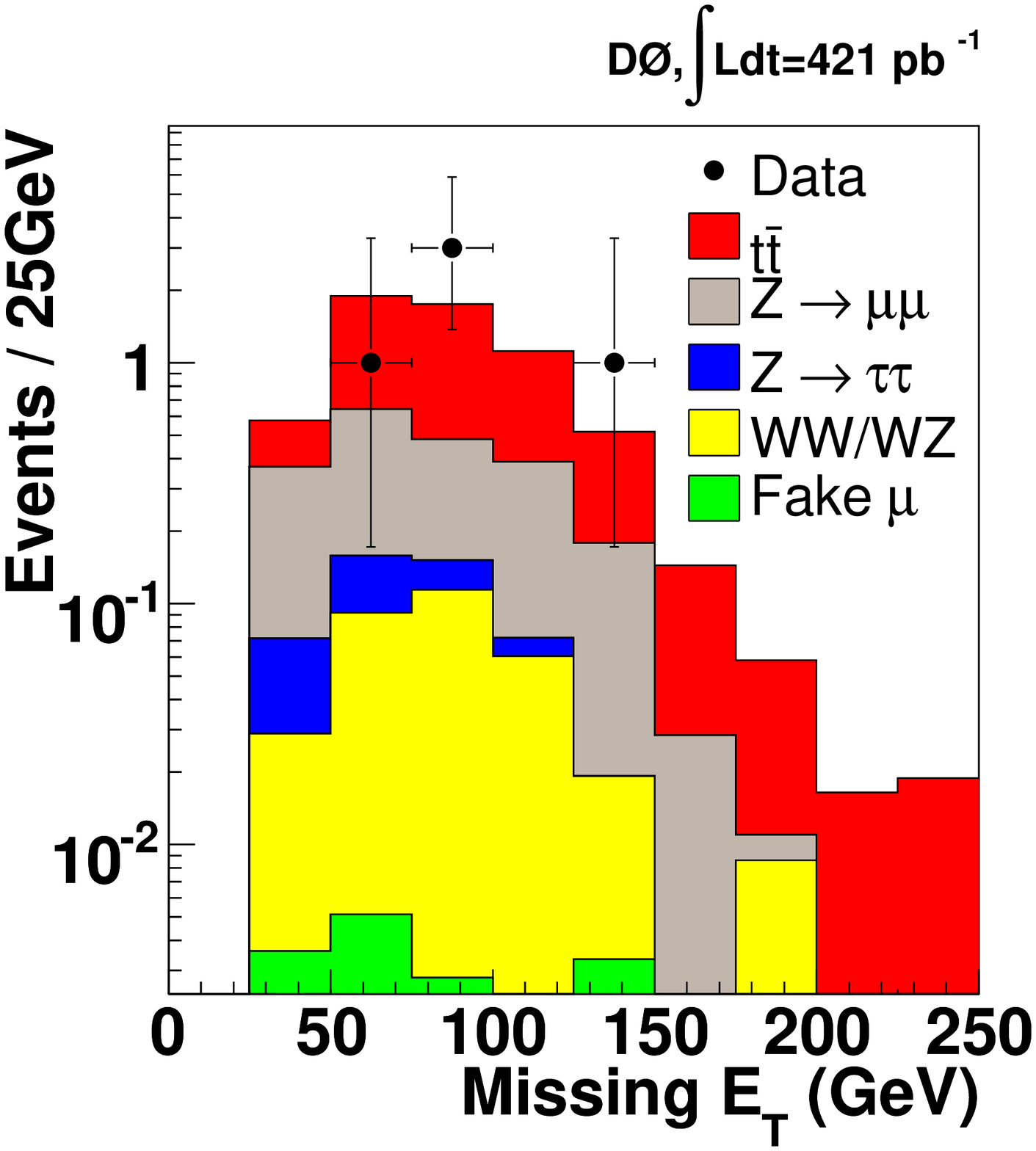}
\caption{\label{fig:mumu_met} Expected and observed \met\ distributions
in dimuon events before (left) and after (right) the two-dimensional cut
in the $\Delta \phi(\mu_{\text{leading}},\met)$ versus \met\ plane.}
\end{figure*}

Candidate events were required to contain two loose, isolated muons
(as described in Sec.~\ref{sec:muID}) with matching central tracks of
opposite charge and two jets with $p_T>20$ GeV.
The first line of Table~\ref{tab:mumu_yields} shows signal and
background yields after this initial event selection. The data
selected are heavily dominated by misidentified Drell-Yan events. Two
additional selection requirements were designed to specifically target
this background. The first was a contour cut made in the plane formed by
event $\met$ and the opening angle in $\phi$ between the leading $p_T$
muon and the $\met$.  Correlations between these two variables are
caused by the misreconstruction of central tracks matched to muons.  An
event having \met\ less than 45 GeV was immediately rejected.  This cut
was further tightened at low and high values of
$\Delta\phi(\mu_{\text{leading}},\met)$ to $\met>90$ GeV and $\met>95$ GeV
respectively.  Events with
$\Delta\phi(\mu_{\text{leading}},\met)>175^{\circ}$ were removed. As can
be seen in Fig.~\ref{fig:mumu_met} and the second line of
Table~\ref{tab:mumu_yields}, the contour cut effectively suppressed the
misidentified background.

\begin{figure}
\includegraphics[scale=0.35]{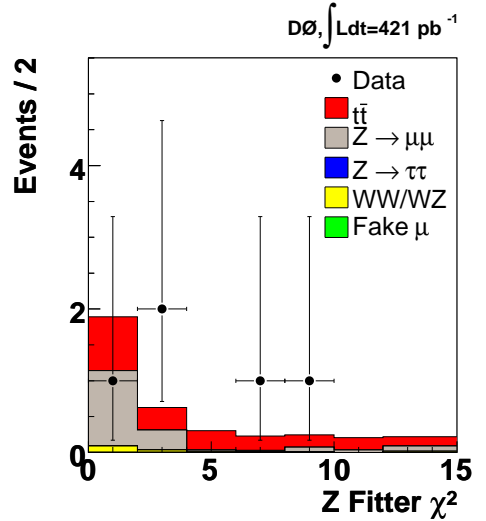}
\caption{\label{fig:chisq} The observed and predicted $\chi^2$
distributions of  dimuon events before the $\chi^2$ cut.} 
\end{figure}

Further background rejection was achieved by cutting on the
compatibility of an event with the $Z \rightarrow \mu\mu$ hypothesis.
To this end, a $\chi^2$ was formed using a $Z$ boson mass constraint and
the measured muon momentum resolution.  The resulting variable, shown in
Fig.~\ref{fig:chisq}, accounts for the $p_T$ and $\eta$ dependence of
the tracking resolution and was found to perform better than a simple
dimuon mass cut.  The final cut value ($\chi^2>4$) and the location and
shape of the contour cut described above were chosen using a grid search
over Monte Carlo predictions of signal and background yields.  Both cuts
were varied simultaneously in the search and the best combination was
chosen using a figure of merit related to the expected statistical
significance of the background subtracted signal and including the
expected uncertainty on the $Z+2$ jets background prediction.

After all cuts, two events remain in the data.
Table~\ref{tab:mumu_expected} shows the expected background and  
signal (assuming $m_t=175$ GeV and $\sigma_{t\bar{t}}=7$ pb)
contributions to the final sample.  

\begin{table}
\caption{\label{tab:mumu_expected} 
  A more detailed listing of the expected $\mu\mu$ signal and background yields presented on the last line of Table~\ref{tab:mumu_yields}.
  The expected number of $t\bar{t}$ events was calculated
  assuming $m_t=175$ GeV and $\sigma_{t\bar{t}}=7$ pb.  Uncertainties
  include statistical and systematic contributions added in quadrature.}
\begin{ruledtabular}
\begin{tabular}{l c}
Process $\hspace{22pt}$ Expected number of $\mu\mu+X$ events & \\
\hline
$t\bar{t}$ (MC)            & $2.96^{+0.31}_{-0.35}$  \\
$WW$/$WZ\to \mu\mu+X$ (MC)     & $0.19^{+0.10}_{-0.07}$\\
$Z\to\mu\mu$/$\tau\tau$($\to \mu\mu$)$+X$ (MC)  & $0.47^{+0.17}_{-0.18}$ \\
Isolation fakes	(data)     & $0.01^{+0.01}_{-0.01}$ \\
Total bkg                  & $0.67^{+0.24}_{-0.22}$ \\
\end{tabular}
\end{ruledtabular}
\end{table}

Background estimates for $Z\rightarrow\mu\mu$, $Z\rightarrow\tau\tau
\rightarrow\mu\mu +X$, and $WW$/$WZ$ production were calculated using the
Monte Carlo samples discussed in Sec.~\ref{sec:simulation}.
The background from fake isolated muons was estimated using a procedure
requiring two samples of events:  a ``tight'' sample containing $N_T$
events passing all the dimuon selection criteria and a ``loose'' sample
of $N_L$ events for which only one muon was required to pass isolation
cuts.  These event counts are related to the numbers of events with
signal-like muons ($N_{sl}$) and events with background-like muons 
($N_{bl}$) via the relations: 
\begin{equation}\label{eq:mumu_fake1}
  N_{L} = N_{sl} + N_{bl}
\end{equation}
and
\begin{equation}\label{eq:mumu_fake2}
  N_{T} =  \epsilon_{\text{sig}}N_{sl} + f_{\mu}N_{bl}.
\end{equation}
Here $\epsilon_{\text{sig}}$ is the probability for signal-like muons to pass
isolation requirements and $f_{\mu}$ is the probability for muons in
background events to pass isolation requirements.  Equations
\ref{eq:mumu_fake1} and \ref{eq:mumu_fake2} can be solved for the
falsely isolated muon background in the fully selected sample ({\em
i.e.}, $f_{\mu}N_{bl}$).

The faking probability was estimated as the isolation efficiency for the
second highest $p_T$ muon in dimuon events.  To eliminate bias from $W$ and
$Z$ boson decays, events in the sample were also required to have non-isolated
leading muons and dimuon masses less than 70 GeV.  The isolation
efficiency for  signal-like muons was estimated with {\sc alpgen}
$t\bar{t}\rightarrow \mu\mu +X$ Monte Carlo. 

\subsection{The $\ell+$track Channel}
\label{sec:ltrack}
The final state in the $\ell+$track channel is characterized by two
oppositely charged high $p_T$ leptons, one explicitly reconstructed
as an electron or a muon and the other identified as an isolated track.
Requiring an isolated track as opposed to a fully reconstructed lepton 
allows the recovery of some events lost due to lepton reconstruction
inefficiency and also adds a small number of events with a 
hadronically decaying tau lepton to the sample (5\%/4\% of the 
$e$+track/$\mu$+track data).
Events were also required to contain two high $p_T$ jets 
and a large $\met$. 
The dominant background in this channel originates from Drell-Yan
production of lepton pairs ($Z/\gamma^{*}\rightarrow ee$ or $Z/\gamma^*
\rightarrow \mu\mu$) where misreconstructed objects, resolution effects
or noise can give rise to mismeasured \met.
Additional instrumental backgrounds arise from multijet and
$W$+jets events, where a jet can be misidentified as an isolated lepton
or track.  Sources of irreducible (physics) backgrounds are
$Z/\gamma^*\rightarrow\tau\tau$ and $WW$/$WZ$ production in association
with jets.

Selected events were required to have one tight, isolated lepton
(electron or muon) with $p_T>15$ GeV and an oppositely charged isolated
track with $p_T>15$ GeV.  Neither the lepton nor the isolated track was
allowed to be found within the cone of a reconstructed jet. In addition,
the lepton and the isolated track were required to be separated in
$\Delta \mathcal{R}$ by requiring $\Delta \mathcal{R}({\rm lepton,\
track})>0.5$.

A track was considered isolated if $\mathcal{E}_{\text{halo}}^{\text{trk}}$, defined in Eq.~\ref{eq:muon_trkiso}, was 
less than 0.12. The quality criteria applied to the isolated track were identical to those applied to the 
central muon tracks, with the exception of the $d_{CA}$ significance requirement, where
$\sigma_{d_{CA}}$/$d_{CA} <$ 5 was used.

The candidate events were further required to have two or more jets, each
with $p_T>20$ GeV.  The sample selected by the above requirements is
dominated by $Z/\gamma^{*}$ background events.
This background was partly removed by requirements on the \met\ in each
event.  For the $\mu+$track channel, $\met >35$ GeV ($>25$ GeV) was required
for 70 GeV$\le M_{\mu,\text{trk}}\le$ 110 GeV ($M_{\mu,\text{trk}}<$ 70  
GeV or $M_{\mu,\text{trk}}>110$ GeV). The corresponding cuts in the $e+$track 
channel were $\met > 20$ GeV ($>15$ GeV) for 70 GeV $\le M_{e,\text{trk}}\le$
100 GeV ($M_{e,\text{trk}}<$ 70 GeV or $M_{e,\text{trk}}> 100$ GeV).  The tighter
requirements for the $\mu+$track channel reflect the impact of the
muon-matched central track $p_T$ resolution on the reconstructed
transverse energy. 

\begin{figure*}
\includegraphics[scale=0.35]{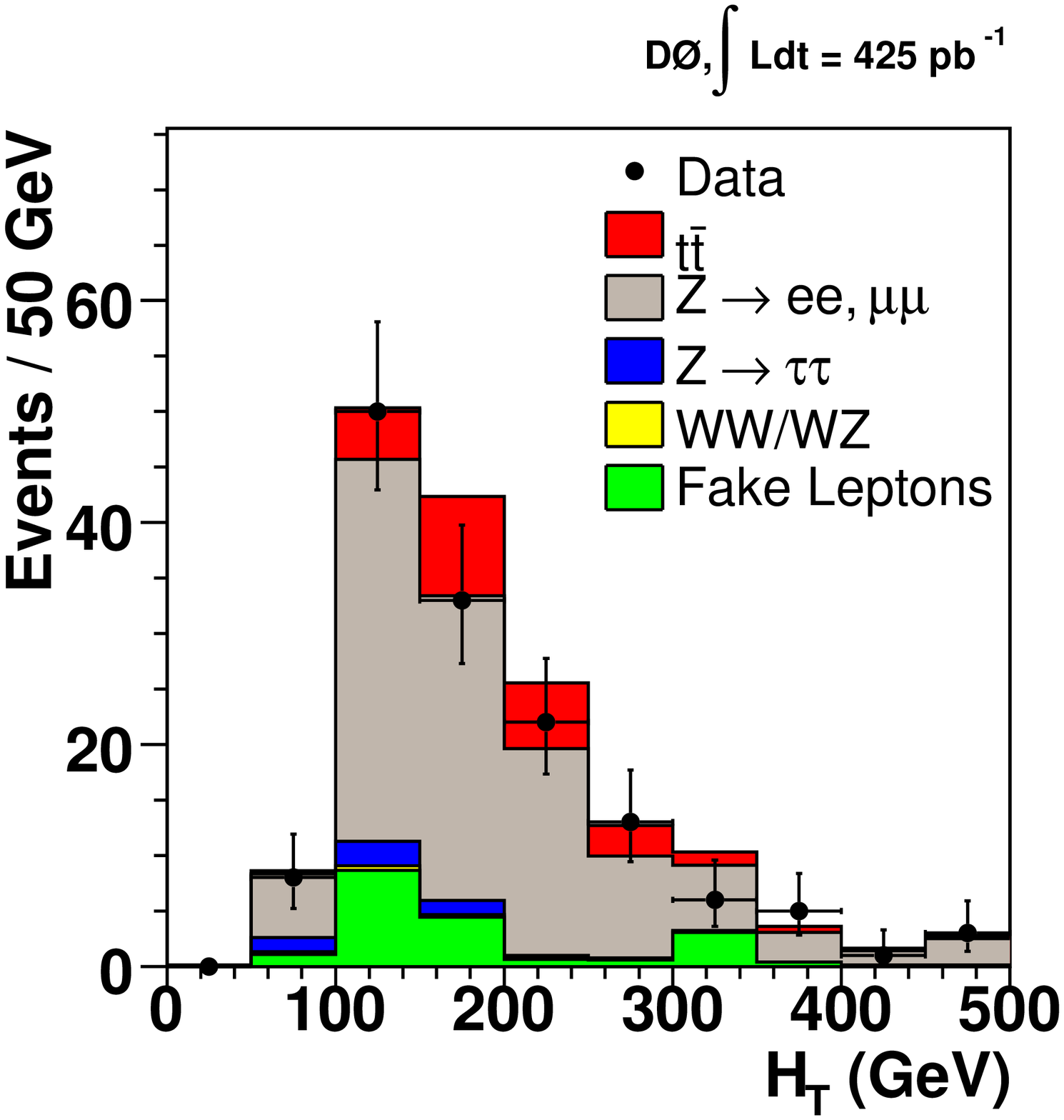}
\includegraphics[scale=0.35]{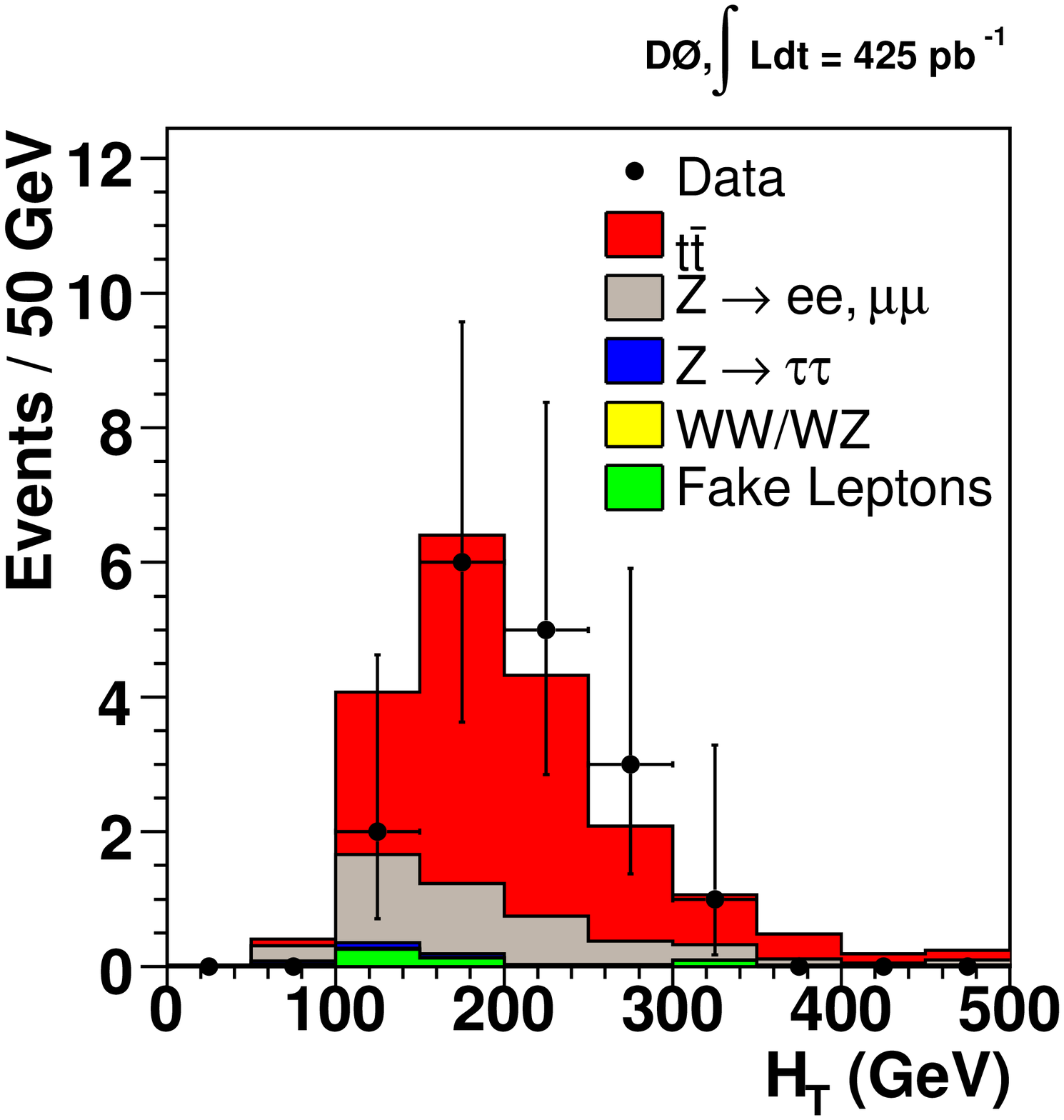}
\caption{\label{fig:ltrackHT} $H_T$ spectra of observed and
predicted events in the $\ell$+track channel before (left) and after
(right) the secondary vertex requirement.}
\end{figure*}

While the preceding requirements are effective, the most powerful
discriminant used to suppress backgrounds in this analysis is the
requirement of at least one $b$ tagged jet, since jets in background
events originate predominantly from light ($u$, $d$, or $s$) quarks or
gluons.  The $b$ tagging algorithm and its performance are discussed in
Sec.~\ref{sec:btag}. Figure~\ref{fig:ltrackHT} shows the numbers of
observed and predicted events with two or more jets as a function of
$H_T$ (defined in Eq.~\ref{eqn:h_t}) before and after the secondary
vertex requirement.  The impact of the tagging requirement on the
backgrounds is clearly visible.  Tables~\ref{tab:etrack_yields} and
\ref{tab:mutrack_yields} show the impact of the \met\ and $b$ tagging
requirements on the expected signal and background yields.

\begin{table*}
\caption{\label{tab:etrack_yields}Numbers of observed and expected
$e+$track events passing the analysis cuts.  The
  instrumental background includes events containing misidentified
  electrons and misreconstructed \met.  Expected
  numbers of $t\bar{t}$ events are for $m_t=175$ GeV and
  $\sigma_{t\bar{t}}=7$ pb.  Uncertainties correspond to statistical and
  systematic contributions added in quadrature.} 
\begin{ruledtabular}
\begin{tabular}{l c c c c c}
&      & Total & Instrumental & Physics & \\
& Data & sig+bkg & bkg  & bkg & $t\bar{t}$ \\ 
\hline
Trigger, $e$+track, $p_T^{e/{\text trk}} > 15$~GeV  & & & & & \\
\quad + $\ge$2 jets, $p_T^{\text{jet}}>20$ GeV
& 436 & $442.2^{+79.9}_{-72.1}$ & $422.6^{+79.5}_{-71.3}$ & $4.7^{+0.9}_{-1.1}$ & $15.0^{+1.0}_{-0.9}$  \\

+ $\met$ cut
& 85 & $92.2^{+20.8}_{-13.8}$ & $74.5^{+20.5}_{-13.2}$ & $3.8^{+0.7}_{-0.9}$ & $13.8^{+1.0}_{-0.9}$  \\

+ $\geq$ 1 $b$ tagged jet
& 11 & $10.9^{+1.2}_{-1.0}$ & $2.7^{+0.9}_{-0.7}$ & $0.1^{+0.0}_{-0.0}$ & $8.1^{+0.7}_{-0.6}$  \\
\end{tabular}
\end{ruledtabular}
\end{table*}

\begin{table*}
\caption{\label{tab:mutrack_yields}Numbers of observed and expected
$\mu+$track events passing the analysis cuts.  The
  instrumental background includes events containing fake isolated muons and
  misreconstructed \met.  Expected numbers of $t\bar{t}$ events are for
  $m_t=175$ GeV and $\sigma_{t\bar{t}}=7$ pb.  Uncertainties correspond
  to statistical and systematic contributions added in quadrature.} 
\begin{ruledtabular}
\begin{tabular}{l c c c c c}
&      & Total & Instrumental & Physics & \\
& Data & sig+bkg & bkg  & bkg & $t\bar{t}$ \\ 
\hline
Trigger, $\mu$+track, $p_T^{\mu/{\text trk}} > 15$~GeV  & & & & & \\
\quad + $\ge$2 jets, $p_T^{\text{jet}}>20$ GeV
& 480 & $483.5^{+89.4}_{-82.8}$ & $465.8^{+88.7}_{-82.0}$ & $4.5^{+0.9}_{-1.1}$ & $13.1^{+1.0}_{-1.0}$  \\

+ $\met$ cut
& 56 & $63.8^{+10.1}_{-10.1}$ & $50.4^{+9.6}_{-9.5}$ & $2.6^{+0.7}_{-0.9}$ & $10.8^{+0.9}_{-0.9}$  \\

+ $\geq$ 1 $b$ tagged jet
& 6 & $8.3^{+0.8}_{-0.8}$ & $1.9^{+0.6}_{-0.6}$ & $0.1^{+0.0}_{-0.0}$ & $6.3^{+0.6}_{-0.6}$  \\
\end{tabular}
\end{ruledtabular}
\end{table*}

After application of all selection criteria, 17 events remain in the data. 
Tables~\ref{tab:etrack_expected} and ~\ref{tab:mutrack_expected} show
the expected signal and background contributions to the final sample.

\begin{table}
\caption{\label{tab:etrack_expected} 
  A more detailed listing of the expected $e$+track signal and background yields presented on the last line of Table~\ref{tab:etrack_yields}.
  The expected number of $t\bar{t}$ events is
  calculated assuming $m_t=175$ GeV and $\sigma_{t\bar{t}}=7$ pb.
  Uncertainties include statistical and systematic contributions added
  in quadrature.}
\begin{ruledtabular}
\begin{tabular}{l c}
Process & Expected number of $e$+track events \\
\hline
$t\bar{t}$ (MC)        & $8.08^{+0.08}_{-0.08}$ \\
$WW$ (MC)              & $0.02^{+1.09}_{-1.09}$ \\
$Z \to ee,\mu\mu$ (MC) & $2.29^{+0.40}_{-0.31}$ \\
$Z \to \tau\tau$  (MC) & $0.12^{+0.31}_{-0.35}$ \\
$W$/multijet (data)    & $0.42^{+0.37}_{-0.37}$ \\
Total bkg              & $2.85^{+0.33}_{-0.27}$ \\
\end{tabular}
\end{ruledtabular}
\end{table}

\begin{table}
\caption{\label{tab:mutrack_expected} 
  A more detailed listing of the expected $\mu$+track signal and background yields presented on the last line of Table~\ref{tab:mutrack_yields}.
  The expected number of $t\bar{t}$ events is
  calculated assuming $m_t=175$ GeV and $\sigma_{t\bar{t}}=7$ pb.
  Uncertainties include statistical and systematic contributions added
  in quadrature.}
\begin{ruledtabular}
\begin{tabular}{l c}
Process & Expected number of $\mu$+track events \\
\hline
$t\bar{t}$ (MC)        & $6.29^{+0.09}_{-0.09}$ \\
$WW$ (MC)              & $0.01^{+1.10}_{-1.12}$ \\
$Z \to ee,\mu\mu$ (MC) & $1.83^{+0.31}_{-0.30}$ \\
$Z \to \tau\tau$ (MC)  & $0.08^{+0.36}_{-0.46}$ \\
$W$/multijet (data)    & $0.08^{+0.88}_{-0.88}$ \\
Total bkg              & $2.00^{+0.29}_{-0.30}$ \\
\end{tabular}
\end{ruledtabular}
\end{table}

The $b$ quark and $c$ quark tagging efficiencies and mistag rate are
parameterized as functions of jet $p_T$ and $|\eta|$ (see
Sec.~\ref{sec:btag}).  The tagging probabilities for $t\bar{t}$ events
were estimated by applying these parameterizations to jets in
simulated events.

The $Z/\gamma^{*}\rightarrow ee/\mu\mu$,
$Z/\gamma^*\rightarrow\tau\tau$, and diboson backgrounds were estimated 
using the simulated samples discussed in Sec.~\ref{sec:simulation}. The
$b$ tagging probability for $Z/\gamma^{*}\rightarrow ee/\mu\mu$
events was estimated using a control sample selected with all
$\ell+$track event selection criteria with the exception of the $\met$ 
cut, which was reversed.  As a cross check, separate $b$ tagging
probabilities were measured for the $e+$track and $\mu+$track channels
and were found to be consistent. The $b$ tagging efficiency for
$Z/\gamma^*\rightarrow\tau\tau$ was taken to be the same as for 
$Z/\gamma^{*}\rightarrow ee,\mu\mu$ events. For diboson events, 
the $b$ tagging efficiency was assumed to be the same as for $W$ boson events 
with associated jet production.

The contribution from $W$ boson and multijet events containing fake
isolated leptons and/or tracks was estimated using specially selected
data samples.  The method employed is similar to that described for
similar backgrounds in the dimuon analysis (Sec.~\ref{sec:mumu}).
Because the effects of loosening isolation requirements on leptons and
tracks were included separately, the system of equations \ref{eq:mumu_fake1}
and \ref{eq:mumu_fake2} expands to include four equations.  The unknowns in
each expression include the number of events with fake 
isolated tracks and fake isolated leptons ($N^{ft}_{fl}$), the event
count with fake isolated tracks and real isolated leptons
($N^{ft}_{rl}$), the number of events with real isolated tracks and fake
isolated leptons ($N^{rt}_{fl}$), and the event count with real isolated
tracks and real isolated leptons ($N^{rt}_{rl}$).  Once these quantities
are obtained, backgrounds due to fake isolated leptons and/or tracks
can be computed as 
\begin{equation}\label{eq:ltrack_fake1}
  N_{W+\text{jets}} = N^{ft}_{rl} + N^{rt}_{fl}
\end{equation}
and 
\begin{equation}\label{eq:ltrack_fake2}
  N_{\text{QCD}} = N^{ft}_{fl}.
\end{equation}
In order to limit the impact of statistical fluctuations, this procedure
was applied to events selected without $b$ tagging.  The final
background estimates were then derived using separate tagging
efficiencies for $W$+jets and multijet backgrounds.

Knowledge of the efficiencies of the tight track and lepton criteria
relative to their loose counterparts for both signal-like and
background-like objects is required to extract the unknown yields from
observed event counts.  The efficiency for a signal-like loose electron,
muon or track to pass each corresponding tight criterion was determined
from simulated samples of $Z/\gamma^{*}\rightarrow ee$ and $\mu\mu$
events.  The efficiencies for loose fake leptons were measured in a 
multijet data sample obtained by selecting lepton events in a low \met\
($<$10 GeV) region. Biases from $Z$ boson decays were removed by
eliminating events with two like-flavor leptons or with an additional
isolated track that, when paired with the lepton, formed an invariant
mass consistent with $M_Z$.  Efficiencies for loose fake tracks were
measured in similarly chosen samples.  

Secondary vertex tagging efficiencies for $W$+jets events were measured
using single-lepton events selected with the same $Z$ boson rejection
criteria used for the multijet sample described above. Additional
biases from the presence of top quark pairs were accounted for by 
subtracting predicted $t\bar{t}$ contributions calculated with an
assumed production cross section of 7 pb.  The event tagging 
probabilities for multijet events were determined using the same 
multijet data samples used to estimate isolation faking probabilities.

\subsection{Summary}
\begin{table}
\caption{\label{tab:yield_summary} Numbers of observed events, expected
  background yields, the product of $t\bar{t}$ selection efficiency
  times branching ratio, and the integrated luminosity for each analysis
  channel. The branching fractions for the $e\mu$, $e e$ and $\mu\mu$
  channels considers the decays $t\bar{t}\to b\bar{b}WW\to e\mu/ee/\mu\mu+X$
  respectively. Both $e+$track and $\mu+$track channels consider 
  $t\bar{t}\to b\bar{b}WW\to \ell \ell+X$ decays ($\ell=e,\mu,\tau$) with
  the $\tau$ leptons decaying both leptonically and hadronically.}
\begin{ruledtabular}
\begin{tabular}{l c c c c}
Channel & $N_{\text{obs}}$ & $N_{\text{bkg}}$ & $\epsilon \times {\mathcal B}$ (\%) & $\int{\mathcal L}dt$ (pb$^{-1}$) \\
\hline
$e\mu$      & 21 & $4.58^{+2.56}_{-1.77}$ & $0.44\pm0.04$ & $427\pm26$ \\
$e e$       &  5 & $1.12^{+0.22}_{-0.27}$ & $0.13\pm0.02$ & $446\pm27$ \\
$\mu\mu$    &  2 & $0.67^{+0.24}_{-0.22}$ & $0.10\pm0.02$ & $421\pm26$ \\
$e+$track   & 11 & $2.85^{+0.33}_{-0.27}$ & $0.27\pm0.02$ & $425\pm26$ \\
$\mu+$track &  6 & $2.00^{+0.29}_{-0.30}$ & $0.21\pm0.02$ & $422\pm26$ \\
\end{tabular}
\end{ruledtabular}
\end{table}

Table~\ref{tab:yield_summary} presents a summary of event counts
observed in data, expected background yields, the products of $t\bar{t}$
selection efficiencies and branching ratios, and luminosities for each of the
five dilepton analysis channels.  These quantities enter into the
top quark pair production cross section calculations discussed in
Secs.~\ref{sec:individual_xsec} and \ref{sec:combined_xsec}.

\section{Cross Section Calculation}
\label{sec:xsec}
\subsection{Individual Channel Cross Sections}
\label{sec:individual_xsec}
To estimate the production cross section $\sigma_j$ for an individual dilepton channel
$j$, the following likelihood function was defined: 
\begin{eqnarray}
\label{eq:pois}
{\boldmath{L}}(\sigma_j, \{N_j^{\text{obs}}, N_j^{\text{bkg}}, \varepsilon_j \times
{\mathcal B}_j, {\cal{L}}_j \} ) & = & {\cal{P}}(N_j^{\text{obs}},n_j)
\nonumber \\*
& = & \frac{n_j^{N_j^{\text{obs}}}}{N_j^{\text{obs}} !}\, e^{-n_j}, 
\end{eqnarray}
where ${\cal{P}}(N_j^{\text{obs}},n_j)$ is the Poisson probability to observe
$N_j^{\text{obs}}$ events given an expected combined signal and background yield
of
\begin{eqnarray}
n_j & = & \sigma_j\, (\varepsilon_j \times {\mathcal B}_j)\,
   {\cal{L}}_j + N_j^{\text{bkg}}.
\end{eqnarray}
Here ${\cal{L}}_j$ is the luminosity, $\epsilon_j \times {\mathcal B}_j$ is
the product of selection efficiency and branching fraction, and
$N_j^{\text{bkg}}$ is the expected background for channel $j$ (see
Table~\ref{tab:yield_summary}).  The cross section 
is then extracted by minimizing the negative log-likelihood function,

\begin{equation}
-\log {\boldmath L}(\sigma_j, \{N_j^{\text{obs}}, N_j^{\text{bkg}}, \varepsilon_j
\times{\mathcal B}_j, {\cal{L}}_j \} ). 
\end{equation}
The results are presented in Table~\ref{tab:ind_xsecs}.

\begin{table}
\caption{\label{tab:ind_xsecs} The $t\bar{t}$ production cross sections
  at $\sqrt{s} = 1.96$~TeV and for a top quark mass of 175 GeV as measured
  in each analysis channel.   The first uncertainty listed for each
  result is statistical in origin.  The second uncertainty is the
  combined effect of all systematics, excluding the uncertainty on the
  luminosity measurement.  The final error listed is from the luminosity
  measurement.  The origins of the systematic uncertainties are
  discussed in Sec.~\ref{sec:systematics}.}
\begin{ruledtabular}
\begin{tabular}{l c}
Channel & $\sigma_{t\bar{t}}$ (pb) \\
\hline
$e\mu$      & 8.8 $^{+2.6}_{-2.3}$ $^{+1.4}_{-1.1}$ $\pm$0.5 \\
$e e$       & 6.7 $^{+4.5}_{-3.3}$ $^{+1.1}_{-0.8}$ $\pm$0.4 \\
$\mu\mu$    & 3.1 $^{+4.2}_{-2.6}$ $^{+0.9}_{-0.9}$ $\pm$0.2 \\
$e$+track   & 7.1 $^{+3.2}_{-2.6}$ $^{+1.0}_{-1.2}$ $\pm$0.4 \\
$\mu$+track & 4.5 $^{+3.1}_{-2.4}$ $^{+0.9}_{-0.9}$ $\pm$0.3 \\
\end{tabular}
\end{ruledtabular}
\end{table}

\subsection{Systematic Uncertainties}
\label{sec:systematics}
Systematic uncertainties for the analyses can be broadly grouped into
those related to signal acceptance calculations and those
concerning overall background estimates.  Brief descriptions of the
sources of systematics are provided below.
\begin{itemize}
\item {\em Primary vertex identification} \\*
  A correction to the simulated efficiency for primary vertex
  selection was estimated by comparing its value in $Z\to ee$/$\mu\mu$
  decays in data and Monte Carlo.  In order to quote a systematic
  uncertainty related to this correction, the ratio of data and Monte
  Carlo efficiencies was varied by $\pm$1$\sigma$, and signal
  efficiencies and expected background yields were re-computed.  The
  ultimate origin of the uncertainty $\sigma$ is the statistical
  limitations of the $Z\to ee$/$\mu\mu$ data samples.
\item {\em Lepton identification} \\*
  For electrons and muons, uncertainties related to the identification
  and selection criteria described in Secs.~\ref{sec:muID} and  
  \ref{sec:eID} were estimated using control samples of
  $Z\to ee$/$\mu\mu$ decays.  The tag and probe technique discussed in
  Sec.~\ref{sec:triggers} was used to compute the effects of each
  criterion in data and Monte Carlo, and the ratio of the resulting
  efficiencies was used to correct the simulation.  When a correction was
  found to depend on object kinematics, it was binned appropriately.
  The corresponding systematic uncertainty was computed by varying the
  correction by $\pm$1$\sigma$, where $\sigma$ arose from the
  statistical limitations of the $Z$ boson control samples in data.
\item {\em Track reconstruction} \\*
  Analyses using muons and the $\ell$+track channels include
  uncertainties associated with central track reconstruction.  Chief
  among these is the uncertainty in the track smearing
  procedure discussed in Sec.~\ref{sec:tracks}.  Signal efficiency
  and background yield calculations were repeated with the smearing
  parameters varied according to their uncertainties, whose
  ultimate origin is in the parameterization of the smearing functions and the
  size of the data samples used to tune them. Because of the
  significance of Bremsstrahlung energy loss for electrons, separate
  uncertainties were used for the $e$+track channel.
\item {\em Jet identification} \\*
  This uncertainty corresponds to the correction to simulated jet
  identification and quality requirement efficiencies described in
  Sec.~\ref{sec:jet_qual}. 
\item {\em Jet energy calibration} \\* 
  This uncertainty includes contributions estimated for the jet energy
  scale and resolution corrections described in Sec.~\ref{sec:JES}.
\item {\em Trigger simulation} \\* 
  Uncertainties on the fits to the trigger efficiencies for each 
  object discussed in Sec.~\ref{sec:triggers} were propagated to
  estimate event triggering systematics.
\item {\em Background estimation} \\* 
  For background estimates using Monte Carlo simulations, normalization
  uncertainties were calculated using theoretical and/or experimental
  uncertainties in the corresponding product of production cross section
  and decay branching ratios. For instance, a systematic uncertainty 
  of 35\% 
  is associated to the normalization of diboson$+$jets background,
  taken very conservatively as the difference between LO and NLO cross
  sections. For backgrounds estimated from data,
  systematic uncertainties have their ultimate origin in the statistical
  limitations of the relevant control samples.
\item {\em $t\bar{t}$ tagging probability} \\* 
  For the $\ell$+track channels, additional uncertainties associated
  with the $b$ quark tagging probability in $t\bar{t}$ decays
  (Sec.~\ref{sec:btag}) were included. 
  The dominant sources of uncertainty arise from the method used
  to extract the $b$ tagging efficiency in data and from the limited
  statistics in the heavy flavor enriched data samples.
\item {\em Background tagging probability} \\* 
  For the $\ell$+track channels, uncertainties in the $b$ tagging
  probabilities for the background processes (Sec.~\ref{sec:ltrack})
  were also taken into account.  These originated from limited
  statistics in the background-enriched data samples and observed 
  dependence of the event tagging probability on \met. For the
  $W$+jets events, where the $t\bar{t}$ contamination was subtracted
  assuming a cross section of 7~pb, the effect of varying the $t\bar{t}$
  cross section between 5 and 9~pb was propagated to the final result.
\item {\em Luminosity} \\* 
  The integrated luminosity corresponding to each of the data samples
  used by the analyses has a fractional uncertainty of
  6.1\%~\cite{Andeen:2007}.
\item {\em Other uncertainties} \\*
  Statistical uncertainties related to the sizes of Monte Carlo and data
  samples used independently for each channel are uncorrelated between
  them.  These and other less important uncertainties are combined in
  this category.
\end{itemize}

For each channel, an uncertainty on the cross section was obtained for
each independent source of systematic uncertainty by varying the source
by $\pm1\sigma$ and propagating the variation into
both background estimates and the signal efficiency. A new likelihood
function was derived for each such variation to give a new optimal cross
section.  The resulting variations in the central value of the cross
section are presented in Table~\ref{tab:ind_systematics}.

\begin{table*}
\caption{\label{tab:ind_systematics} A summary of the effects of
  individual systematic uncertainties on each channel's measured cross 
  section.  Quantities are presented in percent change from the central
  values presented in Table~\ref{tab:ind_xsecs}.}
\begin{ruledtabular}
\begin{tabular}{l | c c | c c | c c | c c | c c}
Source & \multicolumn{2}{c|}{$e\mu$} & \multicolumn{2}{c|}{$ee$} & \multicolumn{2}{c|}{$\mu\mu$} & \multicolumn{2}{c|}{$e$+track} & \multicolumn{2}{c}{$\mu$+track} \\
\hline
PV identification      &  0.5 &  $-0.5$  &  0.5 &  $-0.5$ &  0.7 & $ -0.5$ &  0.6 & $ -0.6$ &  0.7 & $ -0.7$ \\
Lepton identification  &  6.3 &  $-5.9$  &  9.1 &  $-8.2$ &  3.8 & $ -3.5$ &  4.3 & $ -4.2$ &  7.8 & $ -7.2$ \\
Track reconstruction   &  4.2 &  $-4.6$  &  0.1 &  $-0.1$ &  8.8 & $-10.3$ &  3.4 & $ -3.3$ &  4.5 & $ -4.0$ \\
Jet identification     &  4.8 &  $-3.5$  &  8.0 &  $-4.0$ & 13.2 & $ -8.6$ &  2.8 & $ -6.3$ &  6.6 & $ -0.8$ \\
Jet energy calibration &  7.1 &  $-6.1$  &  8.2 &  $-5.4$ & 22.0 & $-21.7$ &  8.5 & $-11.2$ & 10.0 & $-11.4$ \\
Trigger                & 10.2 &  $-5.7$  &  7.5 &  $-1.9$ &  5.5 & $ -4.0$ &  2.9 & $ -2.3$ &  5.5 & $ -4.4$ \\
Bkg estimation         &  4.7 &  $-3.7$  &  2.1 &  $-2.2$ &  5.6 & $ -5.4$ &  2.5 & $ -2.5$ &  3.8 & $ -3.9$ \\
$t\bar{t}$ tagging      &  0.0 &  $-0.0$  &  0.0 &  $-0.0$ &  0.0 & $ -0.0$ &  4.0 & $ -3.8$ &  4.0 & $ -3.8$ \\
Bkg tagging            &  0.0 &  $-0.0$  &  0.0 &  $-0.0$ &  0.0 & $ -0.0$ &  7.0 & $ -7.0$ & 11.2 & $-11.2$ \\
Other                  &  2.4 &  $-2.3$  &  2.3 &  $-2.2$ &  2.6 & $ -2.4$ &  2.2 & $ -2.3$ &  2.6 & $ -2.7$ \\
\hline
Total                  & 16.2 &  $-12.5$ & 16.7 & $-11.2$ & 28.6 & $-26.7$ & 13.9 & $-16.5$ & 20.5 & $-19.6$ \\
\end{tabular}
\end{ruledtabular}
\end{table*}

\subsection{Combined Cross Section}
\label{sec:combined_xsec}
Calculation of the combined estimate of the $t\bar{t}$ production cross
section using all of the results presented in Table~\ref{tab:ind_xsecs}
is complicated by the fact that some of the selection criteria are
correlated.  Specifically, the $ee$ criteria are correlated with those
of the $e$+track analysis, and the $\mu\mu$ and $\mu$+track criteria 
overlap.  To account for this, we apply a BLUE technique ({\em i.e.},
Best Linear Unbiased Estimate)~\cite{Lyons:1988}.

The correlations between the top quark pair selection efficiencies of the
non-orthogonal analysis pairs were estimated with psuedo-experiments
drawn from Monte Carlo samples.  The $ee$--$e$+track and
$\mu\mu$--$\mu$+track correlations were found to be 43\% and 47\%,
respectively.  The use of $b$ tagging in the $\ell$+track selections
resulted in negligible correlations between the backgrounds surviving
each channel's selection.

Correlations between the systematic uncertainties of each of the
analyses were also included in the combination.
These were taken as 100\%, $-100$\%, or 0\%, as appropriate.  Furthermore,
all asymmetric uncertainties were made symmetric by averaging their
positive and negative values. 
The combined cross section was derived using an iterative process. The
combination in each iteration step was performed using the expected
statistical and systematic uncertainties evaluated at the cross section
obtained in the previous iteration step. The use of expected uncertainties
avoids over-weighting the results of downward fluctuations. It was
verified that the result of the iterative process was independent
of the cross section input to the first iteration.
The calculation was repeated until the result was stable to within
0.01\% between iterations.  The resulting combined cross section is 
\begin{equation}
\sigma_{t\bar{t}} = 7.4 \pm1.4\thinspace{\rm(stat)} \pm0.9\thinspace{\rm(syst)}
\pm0.5\thinspace{\rm(lumi)}\, {\rm pb}
\end{equation}
for a top quark mass of 175 GeV.  There were a total of four degrees of
freedom in the combination, and the $\chi^2$ of the result is 1.6.
Table~\ref{tab:combo_weights} lists the relative weight of each analysis
channel's result in the combination.  The $e\mu$ measurement dominates
the result, with the two $\ell$+track results entering with the next
most significant weights.  Table~\ref{tab:comb_systematics} presents the
contribution of each individual systematic uncertainty described in
Sec.~\ref{sec:systematics} to the total error on the result. 

\begin{table}
\caption{\label{tab:combo_weights} Relative weight of 
each measurement in the combined cross section calculation.}
\begin{ruledtabular}
\begin{tabular}{l c}
Channel & Weight (\%) \\
\hline
$e\mu$      &   53 \\
$e$+track   &   22 \\
$\mu$+track &   17 \\
$ee$        &    4 \\
$\mu\mu$    &    4 \\
\hline	       
All         &  100 \\
\end{tabular}
\end{ruledtabular}
\end{table}

\begin{table}
\caption{\label{tab:comb_systematics} A summary of the effects of
  individual systematic uncertainties on the combined cross 
  section measurement.}
\begin{ruledtabular}
\begin{tabular}{l c}
Source & Uncertainty (pb) \\
\hline
PV identification      & 0.07 \\
Lepton identification  & 0.41 \\
Track reconstruction   & 0.09 \\
Jet identification     & 0.30 \\
Jet energy calibration & 0.60 \\
Trigger                & 0.39 \\
Bkg estimation         & 0.22 \\
$t\bar{t}$ tagging     & 0.11 \\
Bkg tagging            & 0.19 \\
Other                  & 0.10 \\
\hline
Total                  & 0.94 \\
\end{tabular}
\end{ruledtabular}
\end{table}

\begin{figure}
\includegraphics[scale=0.40]{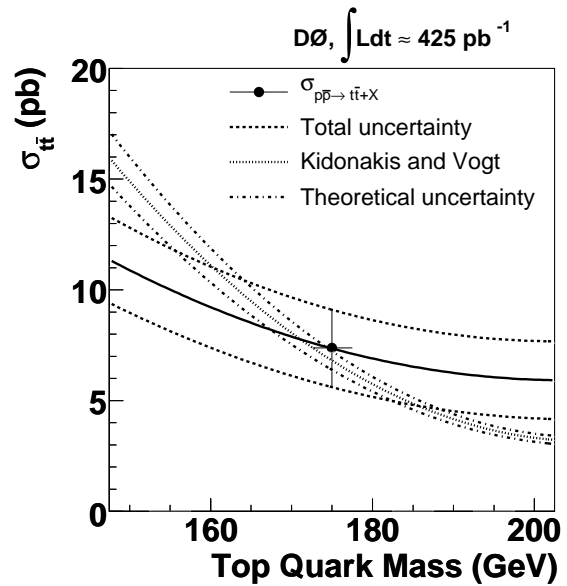}
\caption{\label{fig:massplot} The dependence of the
measured cross section on the top quark mass compared to the theoretical
prediction~\cite{Cacciari:2003fi,Kidonakis:2003qe}.}
\end{figure}

The dependence of the result on the top quark mass was computed using
parameterizations of each channel's selection efficiency as a function of
$m_t$.  For a set of assumed masses, the individual channel results
and their combination were recalculated using the appropriate
efficiency.  The result is shown in Fig.~\ref{fig:massplot}.  For values
of $m_t$ between 170~GeV and 180~GeV, the value of the measured
cross section as a function of top quark mass is approximated by:
\begin{equation}
\label{eq:xsvsmass}
\sigma_{t\bar{t}}(m_t) = 7.4\thinspace{\rm pb} - 0.1 \frac{{\rm pb}}{{\rm GeV}}
\times (m_t - 175\, {\rm GeV}).
\end{equation}

For the current Tevatron average of top quark mass of 
170.9 GeV~\cite{TevEWWG:2007}, the resulting value of the 
cross section is 7.8 $\pm$1.8\thinspace(stat+syst) pb.

\section{Conclusion}
\label{sec:conclusion}
We have measured the $t\bar{t}$ production cross section in $p\bar{p}$
collisions at \( \sqrt{s}=1.96 \) TeV utilizing dilepton signatures
in approximately 425 pb$^{-1}$ of data collected with the D0 detector.
The result, for $m_t$ = 175 GeV, is 
\begin{equation}
\sigma_{t\bar{t}} = 7.4 \pm1.4\thinspace{\rm(stat)} \pm1.0\thinspace{\rm(syst)}\, {\rm pb}.
\end{equation}
This is in good agreement with the theoretical prediction of 
6.7$^{+0.7}_{-0.9}$ pb from the full NLO matrix elements and the
re-summation of the leading and next-to-leading soft
logarithms~\cite{Cacciari:2003fi,Kidonakis:2003qe}. 
For the current Tevatron average of
$m_t$ = 170.9 GeV, the corresponding value of the 
measured cross section is 7.8 $\pm$1.8\thinspace(stat+syst) pb.

\section{Acknowledgements}
%
We thank the staffs at Fermilab and collaborating institutions, 
and acknowledge support from the 
DOE and NSF (USA);
CEA and CNRS/IN2P3 (France);
FASI, Rosatom and RFBR (Russia);
CAPES, CNPq, FAPERJ, FAPESP and FUNDUNESP (Brazil);
DAE and DST (India);
Colciencias (Colombia);
CONACyT (Mexico);
KRF and KOSEF (Korea);
CONICET and UBACyT (Argentina);
FOM (The Netherlands);
Science and Technology Facilities Council (United Kingdom);
MSMT and GACR (Czech Republic);
CRC Program, CFI, NSERC and WestGrid Project (Canada);
BMBF and DFG (Germany);
SFI (Ireland);
The Swedish Research Council (Sweden);
CAS and CNSF (China);
Alexander von Humboldt Foundation;
and the Marie Curie Program.
%

\appendix

\section{Trigger Requirements}
\label{sec:trig_appendix}
Tables~\ref{tab:emu_triggers}--\ref{tab:mutrack_triggers}
list the  trigger conditions used for each analysis channel.  A
description of the triggering system and details regarding particle
reconstruction in each trigger subsystem are available in
Sec.~\ref{sec:triggers}.  As beam
conditions changed and delivered instantaneous luminosity increased,
trigger conditions were changed to maintain event selection rates within 
operational limits.  The tables group triggers used together at the same
time and present the total luminosity exposed to each grouping.

\begin{table*}
\caption{\label{tab:emu_triggers}Trigger requirements used to collect
  data for the $e\mu$ analysis.  Total integrated luminosity exposed to each
  trigger set is given in the last column.}
\begin{ruledtabular}
\begin{tabular}{c c c c}
Level 1 & Level 2 & Level 3 & Integrated \\
conditions & conditions & conditions & luminosity (pb$^{-1}$) \\
\hline
1 $e$, $E_T>5$ GeV and 1 loose $\mu$& none& 1 loose $e$, $E_T>10$ GeV & 130.2 \\
\hline
1 $e$, $E_T>6$ GeV and 1 loose $\mu$& none&1 loose $e$, $E_T>12$ GeV & 243.8  \\
\hline
& & 1 loose $e$, $E_T>12$ GeV & \multirow{5}{*}{53.2} \\
& & OR & \\
1 $e$, $E_T>6$ GeV and 1 loose $\mu$& 1 $\mu$ & 1 tight $e$, $E_T>7$ GeV & \\
& & OR & \\
& & 1 tight, track-matched $e$, $E_T>5$ GeV & \\
\end{tabular}
\end{ruledtabular}
\end{table*}

\begin{table*}
\caption{\label{tab:ee_triggers}Trigger requirements used to collect
  data for the $ee$ analysis.  Total integrated luminosity exposed to each
  trigger set is given in the last column.}
\begin{ruledtabular}
\begin{tabular}{c c c c}
Level 1 & Level 2 & Level 3 & Integrated \\
conditions & conditions & conditions & luminosity (pb$^{-1}$) \\
\hline
2 $e$, $E_T>10$ GeV & none & 1 loose $e$, $E_T>10$ GeV & 23.3  \\
\hline
2 $e$, $E_T>10$ GeV & none & 1 loose $e$, $E_T>20$ GeV & 120.2  \\
\hline
1 $e$, $E_T>11$ GeV & & 2 loose $e$, $E_T>20$ GeV & \multirow{5}{*}{252.2} \\
OR & & OR & \\
2 $e$, $E_T>6$ GeV & none & 1 loose and 1 tight $e$, $E_T>15$ GeV & \\
OR & & & \\
2 $e$, $E_T^1>9$ GeV, $E_T^2>3$ GeV & & & \\
\hline
1 $e$, $E_T>11$ GeV & & 2 loose $e$, $E_T>20$ GeV & \multirow{5}{*}{49.8} \\
OR & & OR & \\
2 $e$, $E_T>6$ GeV & 2 $e$, $\sum E_T>18$ GeV & 2 tight $e$, $E_T>8$ GeV~\footnote{For part of the data, a 10 GeV $E_T$ requirement was used.} & \\
OR & & OR & \\
2 $e$, $E_T^1>9$ GeV, $E_T^2>3$ GeV & & 1 loose and 1 tight $e$, $E_T>15$ GeV & \\
\end{tabular}
\end{ruledtabular}
\end{table*}

\begin{table*}
\caption{\label{tab:mumu_triggers}Trigger requirements used to collect
  data for the $\mu\mu$ analysis.  Each triggering regime has both
  a single muon and a dimuon requirement, each of which ties together
  trigger conditions at all three levels.  Total integrated luminosity
  exposed to each trigger set is given in the last column.}
\begin{ruledtabular}
\begin{tabular}{c c c c c}
Multiplicity & Level 1 & Level 2 & Level 3 & Integrated \\
 & conditions & conditions & conditions & luminosity (pb$^{-1}$) \\
\hline
1 & 1 tight $\mu$ &  1 $\mu$, $p_T>3$ GeV & 1 track, $p_T>10$ GeV&  \multirow{2}{*}{59.5} \\
\cline{2-4}
2 & 2 loose $\mu$ &  1 $\mu$ & none &  \\
\hline
1 & 1 tight $\mu$ &  1 $\mu$, $p_T>3$ GeV & 1 track, $p_T>10$ GeV & \multirow{4}{*}{66.5}  \\
\cline{2-4}
 & 2 loose $\mu$ &  1 $\mu$ & 1 $\mu$, $p_T>$15 GeV & \\
2 & & & OR& \\
 & & & 1 track, $p_T>10$ GeV & \\
\hline
1 & 1 tight $\mu$ &  1 $\mu$, $p_T>3$ GeV & 1 track, $p_T>10$ GeV & \multirow{4}{*}{243.8} \\
\cline{2-4}
 & 2 loose $\mu$ &  1 $\mu$ & 1 $\mu$, $p_T>6$ GeV & \\
2 & & & OR& \\
 & & & 1 track, $p_T>5$ GeV & \\
\hline
1 & 1 tight $\mu$ and 1 track, $p_T>10$ GeV &  1 $\mu$, $p_T>3$ GeV & 1 $\mu$, $p_T>15$ GeV & \multirow{4}{*}{51.5} \\
\cline{2-4}
 & 2 loose $\mu$ &  1 $\mu$ & 1 $\mu$, $p_T>6$ GeV & \\
2 & & & OR& \\
 & & & 1 track, $p_T>5$ GeV & \\
\end{tabular}
\end{ruledtabular}
\end{table*}

\begin{table*}
\caption{\label{tab:etrack_triggers}Trigger requirements used to collect
  data for the $e$+track analysis. 
  For some periods of data collection, a logical OR of multiple
  requirements was used.  Each requirement tied together trigger
  conditions at all three levels.  
  Total integrated luminosity exposed to each trigger set is given in
  the last column.}
\begin{ruledtabular}
  \begin{tabular}{c c c c}
    Level 1 & Level 2 & Level 3 & Integrated \\
    conditions & conditions & conditions & luminosity (pb$^{-1}$) \\

\hline
1 $e$, $E_T > 10$~GeV & 1 $e$, $E_T > 10$~GeV & 1 tight $e$, $E_T > 15$~GeV & \multirow{3}{*}{127.8} \\
AND & AND & AND & \\
2 jets, $E_T > 5$~GeV & 2 jets, $E_T > 10$~GeV & 2 jets, $E_T > 15$~GeV & \\

\hline
1 $e$, $E_T > 11$~GeV & none & 1 tight $e$, $E_T > 15$~GeV& \multirow{13}{*}{244.0} \\
                      & & AND & \\
                      & & 2 jets, $E_T > 20$~GeV & \\
\cline{1-3}
1 $e$, $E_T > 11$~GeV & none & 1 tight $e$, $E_T > 20$~GeV & \\
 OR & & & \\
2 $e$, $E_T > 6$~GeV & & & \\
OR & & & \\
2 $e$, $E_T > 3$~GeV, 1 $e$, $E_T > 9$~GeV & & & \\
\cline{1-3}
1 $e$, $E_T > 11$~GeV & none & 1 loose $e$, $E_T > 50$~GeV & \\
 OR & & & \\
2 $e$, $E_T > 6$~GeV & & & \\
OR & & & \\
2 $e$, $E_T > 3$~GeV, 1 $e$, $E_T > 9$~GeV & & & \\
\hline
1 $e$, $E_T > 11$~GeV & 1 $e$, $E_T > 15$~GeV & 1 tight $e$, $E_T > 15$~GeV& \multirow{15}{*}{53.7} \\
                      & & AND & \\
                      & & 2 jets, $E_T > 20$~GeV & \\
                      & & AND & \\
                      & & 1 jet, $E_T > 25$~GeV & \\
\cline{1-3}
1 $e$, $E_T > 11$~GeV &  1 $e$, $E_T > 15$~GeV & 1 tight $e$, $E_T > 20$~GeV & \\
 OR & & & \\
2 $e$, $E_T > 6$~GeV & & & \\
OR & & & \\
2 $e$, $E_T > 3$~GeV, 1 $e$, $E_T > 9$~GeV & & & \\
\cline{1-3}
1 $e$, $E_T > 11$~GeV & 1 $e$, $E_T > 15$~GeV & 1 loose $e$, $E_T > 50$~GeV & \\
 OR & & & \\
2 $e$, $E_T > 6$~GeV & & & \\
OR & & & \\
2 $e$, $E_T > 3$~GeV, 1 $e$, $E_T > 9$~GeV & & & \\
\hline
  \end{tabular}
\end{ruledtabular}
\end{table*}

\begin{table*}
\caption{\label{tab:mutrack_triggers}Trigger requirements used to collect
  data for the $\mu$+track analysis. 
  For some periods of data collection, a logical OR of multiple
  requirements was used.  Each requirement tied together trigger
  conditions at all three levels.  Total integrated luminosity exposed to
  each trigger set is given in the last column.}
\begin{ruledtabular}
  \begin{tabular}{c c c c}
    Level 1 & Level 2 & Level 3 & Integrated \\
    conditions & conditions & conditions & luminosity (pb$^{-1}$) \\
    \hline
    1 loose $\mu$ & 1 $\mu$ & 1 jet, $E_T > 20$~GeV & \multirow{3}{*}{131.5} \\
    AND & & & \\
    1 jet $E_T > 5$~GeV & & & \\
    \hline
    1 loose $\mu$ & 1 $\mu$ & 1 jet, $E_T > 25$~GeV & \multirow{4}{*}{244.0} \\
    AND & AND & & \\
    1 jet, $E_T > 3$~GeV & 1 jet, $E_T > 10$~GeV & & \\
    \cline{1-3}
    1 tight $\mu$ & 1 $\mu$, $p_T > 3$~GeV & 1 track, $p_T > 10$~GeV &  \\
    \hline
    1 tight $\mu$ & 1 $\mu$ & 1 jet, $E_T > 25$~GeV & \multirow{12}{*}{30.3} \\
    AND & AND & & \\
    1 jet, $E_T > 5$~GeV & 1 jet, $E_T > 8$~GeV & & \\
    \cline{1-3}
    1 loose $\mu$ w. trackmatch & none & 1 track, $p_T > 10$~GeV & \\
    AND & & OR & \\
    1 track, $p_T > 10$~GeV & & 1 $\mu$, $p_T > 15$~GeV & \\
    \cline{1-3}
    1 tight $\mu$ & 1 $\mu$, $p_T > 3$~GeV & 1 $\mu$, $p_T > 15$~GeV & \\
    AND & & & \\
    1 track, $p_T > 10$~GeV & & & \\
    \cline{1-3}
    1 loose $\mu$ & 1 $\mu$ & 1 $\mu$, $p_T > 15$~GeV & \\
    AND & & & \\
    1 track, $p_T > 10$~GeV & & & \\
    \hline    
    1 tight $\mu$ & 1 $\mu$ & 1 $\mu$, $p_T > 3$~GeV & \multirow{12}{*}{16.0} \\
    AND & AND & AND & \\
    1 jet, $E_T > 5$~GeV & 1 jet, $E_T > 8$~GeV & 1 jet, $E_T > 25$~GeV & \\
    \cline{1-3}
    1 loose $\mu$ w. trackmatch & none & 1 track, $p_T > 10$~GeV & \\
    AND & & OR & \\
    1 track, $p_T > 10$~GeV & &  1 $\mu$, $p_T > 15$~GeV & \\
    \cline{1-3}
    1 tight $\mu$ & 1 $\mu$, $p_T > 3$~GeV & 1 $\mu$, $p_T > 15$~GeV & \\
    AND & & & \\
    1 track, $p_T > 10$~GeV & & & \\
    \cline{1-3}
    1 loose $\mu$ & 1 $\mu$ & 1 $\mu$, $p_T > 15$~GeV & \\
    AND & & & \\
    1 track, $p_T > 10$~GeV & & & \\
  \end{tabular}
\end{ruledtabular}
\end{table*}

\end{document}